\documentclass[journal]{IEEEtran}

\usepackage{cite}
\usepackage{amsmath,amssymb,amsfonts}
\usepackage{algorithmic}
\usepackage{graphicx}
\usepackage{textcomp}

\usepackage{siunitx}
\usepackage{subfigure}

\graphicspath{{./pics/}}

\newcommand{\LEFT}{\mbox{~$\circ\!\!-\!\!\!\bullet$~}}

\def\BibTeX{{\rm B\kern-.05em{\sc i\kern-.025em b}\kern-.08em
    T\kern-.1667em\lower.7ex\hbox{E}\kern-.125emX}}

\begin{document}


\title{FSK-based Simultaneous Wireless Information and Power Transfer in Inductively Coupled Resonant Circuits Exploiting Frequency Splitting}

\author{Peter A. Hoeher, Fellow, IEEE

\thanks{Faculty of Engineering, Kiel University, D-24143 Kiel, Germany (e-mail: ph@tf.uni-kiel.de)}}



\maketitle

\begin{abstract}
Inductively coupled resonant circuits are affected by the so-called frequency splitting phenomenon at short distances.
In the area of power electronics, tracking of one of the peak frequencies is state-of-the-art.
In the data transmission community, however, the frequency splitting effect is often ignored.
Particularly, modulation schemes have not yet been adapted to the bifurcation phenomenon.
We argue that binary frequency shift keying (2-ary FSK) is a low-cost modulation scheme which well matches the double-peak
voltage transfer function $H(s)$,
particularly when the quality factor $Q$ is large, whereas most other modulation schemes suffer from the small bandwidths of the peaks.
Additionally we show that a rectified version of 2-ary FSK, coined rectified FSK (RFSK), is even more attractive
from output power and implementation points of view.
Analytical and numerical contributions include the efficiency factor, the impulse response, and the bit error performance.
A low-cost noncoherent receiver is proposed.  Theoretical examinations are supported by an experimental prototype.
\end{abstract}




\section{Introduction}\label{sec1}
Simultaneous wireless information and power transfer (SWIPT) is currently a hot topic.
SWIPT is useful for a wide field of applications, ranging from low-energy harvesting 
(e.g., wireless sensor networks, wearables, and implants), to medium-energy devices 
(e.g., smartphones and notebooks), up to high-power devices (e.g., electric vehicles, robots, and underwater vehicles).  

\begin{figure}[t]
\centering
\includegraphics[scale=0.35]{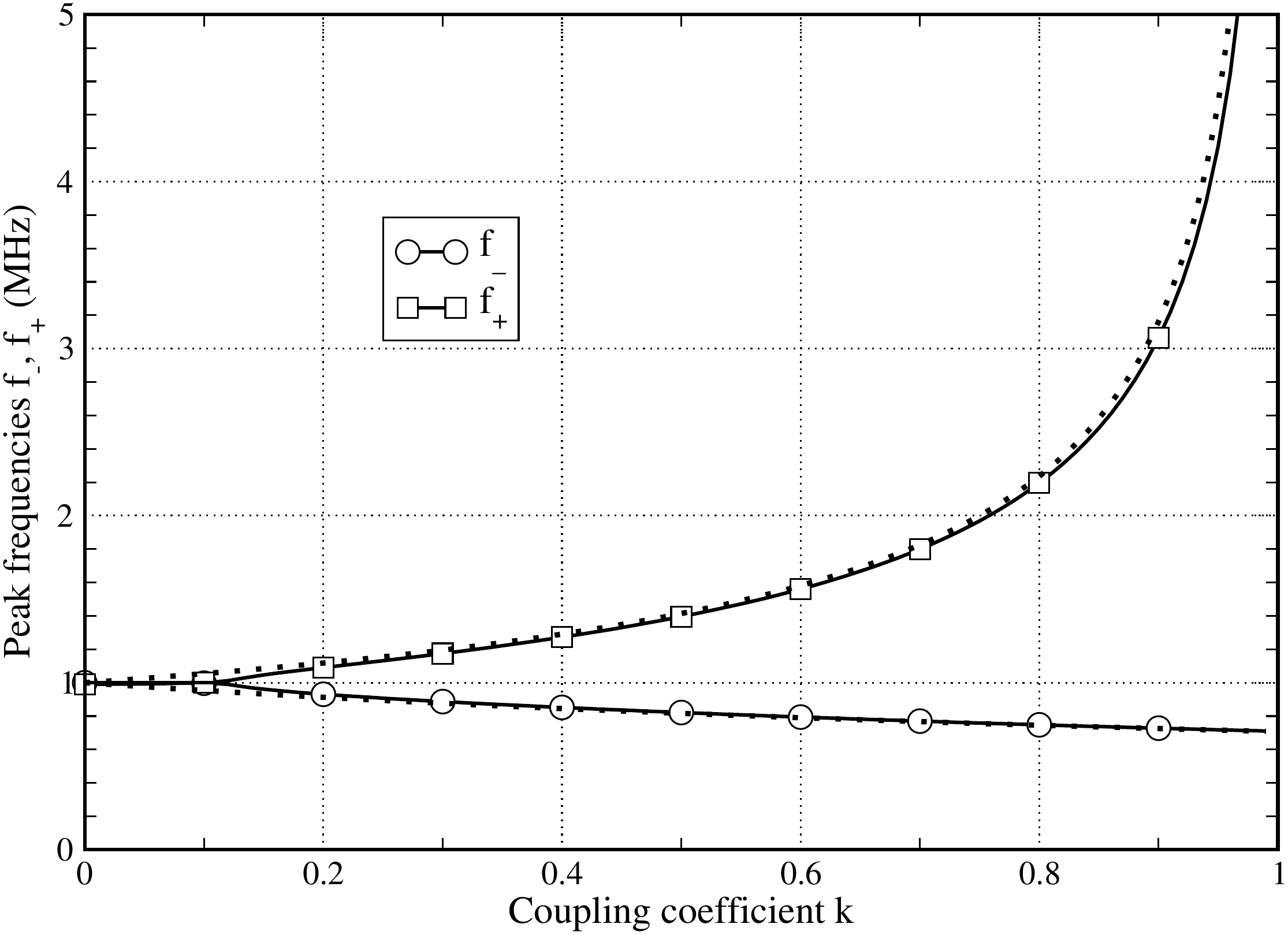}
\caption{Illustration of the frequency splitting phenomenon assuming a two-coil series-resonant circuit given the following parameters:
$C_1=C_2=4$~nF, $L_1=L_2=\SI{6.3}{\micro \henry}$, $R_1=R_2=0.62~\Omega$, $R_S=0.17~\Omega$, $R_L=10~\Omega$.  The meaning of these parameters 
is introduced in Section~\ref{sec2}.  The resonant frequency is at $f_0=\omega_0/(2\pi)=1$~MHz in this example, but the results of this paper can be scaled 
to other frequencies.  Note that at $k=0.6$ we get $f_{+}/f_{-}=2$ and at $k=0.8$ we obtain $f_{+}/f_{-}=3$, 
independent of the remaining system parameters.  Approximation (\ref{approx}) is shown by dotted lines.}
\label{fplus_fminus_vs_k}
\end{figure}

In this article, resonant inductive coupling is assumed between a single primary and a single secondary coil \cite{kur07}.  
Emphasis is on short distances, i.e., on situations with a large coupling coefficient $k$, referred to as tight (or over) coupling.   
If the coupling coefficient $k$ exceeds the so-called splitting coupling coefficient $k_{\mbox{\footnotesize split}}$, 
the voltage transfer function has two peaks denoted 
as $\omega_{-}=2\pi f_{-}$ and $\omega_{+}=2\pi f_{+}$, rather than a single peak at resonant frequency $\omega_0=2\pi f_0$, 
see Fig.~\ref{fplus_fminus_vs_k}.
This effect is known as {\em frequency splitting} or {\em bifurcation} phenomenon \cite{kim07}~--\cite{kim17}.
Frequency splitting focuses on the output characteristic of resonant inductive coupled systems, 
whereas bifurcation on the input characteristic \cite{niu13}.  The relation between these characteristics is very close, however.  
Both with respect to (w.r.t.) power transmission as well as w.r.t.~data transmission 
it is important to consider the bifurcation phenomenon.  

In the area of power electronics, the system is either operated at resonant frequency, or one of the peak frequencies is tracked to enable zero voltage switching, see e.g.~\cite{niu13, zha14, zhe15}.
Subsequently we show that the output power is maximized at the peak frequencies, but the efficiency factor $\eta$ 
is largest at the resonant frequency, disregarding switching and other frequency-dependent losses.  
This is related to the insight that we need to distinguish between maximum power transfer 
and maximum energy efficiency \cite{hui14}.

In the microwave community, the bifurcation phenomenon is used for analog filter realizations \cite{mak01,sun12}.
As opposed to power electronics targeting fairly short ranges, in the field of data transmission larger ranges are frequently of 
interest.  This may explain the observation that most publications on frequency splitting/bifurcation have been issued by the former community, 
just a few by the latter community. One of these exceptions is \cite{ngu15}, where binary chirp modulation is suggested for data 
transmission in inductive MIMO systems.  In the context of SWIPT, however, it is important to investigate modulation schemes that are  
matched to the frequency splitting effect.  The modulation scheme should be designed so that even at short distances
the impact on power transmission is small.  This is a key motivation of this article. In the following, attention is 
on maximizing output power while supporting reasonable data rates. 

We exploit the fact that resonances are narrowband by nature, particularly when $Q$ is large -- 
but due to bifurcation resonances come in pairs. Binary frequency shift keying (2-ary FSK) uses a pair of discrete frequencies, 
and hence can adaptively be matched to bifurcation in a natural way. Binary FSK is not uncommon in SWIPT systems.
Perhaps the most prominent application of 2-ary FSK in wireless power transfer systems is the Qi standard, where the power transmitter
uses 2-ary FSK to provide synchronization and other information to the receiver by modulating its operation frequency \cite{qi17}. 
Contrary to our work, however, both carrier frequencies are fixed in the Qi standard. 
Furthermore, the author is not aware of any in-depth analysis of 2-ary FSK in the presence of frequency splitting.
In the SWIPT literature considering frequency splitting, to our best knowledge the only reference mentioning FSK is \cite{jia15}. 
In \cite{jia15} FSK is noted in a block diagram, but no further motivation and elaboration on FSK is given throughout the text. 
As mentioned in \cite{wu15}, non-adaptive FSK decreases the performance of power transfer when data and power transfer are conducted on the same channel. 
In situations with tight coupling and with a transmitter-side adaptation, however, this drawback can be relieved as shown in this contribution. 

Possible employments of this work are industrial and commercial applications, where power transmission ideally shall not be degraded by 
simultaneous data transmission.  Examples include electric vehicles of any size \cite{cov13}, like cars, robots, and underwater vehicles. 
Another profitable use case are portable electronic products \cite{hui13}.
But also contact-free cable junctions are a potential application when carrying energy plus data.  These near-distance applications are different 
from wireless powered communication networks \cite{bi16, kis16}, low-power medical applications \cite{gho04}, 
radio-frequency identification (RFID) and contactless smartcards \cite{fin10}, often cited in the context of SWIPT.  
In the latter applications, frequency splitting is no issue at medium and long ranges.

Of course there are situations in practice lacking from frequency splitting. Particularly in wireless power charging applications, however, for small coupling coefficients the efficiency is not acceptable -- power transfer should stop anyway.

Own contributions include the following items: 
(i) Binary FSK is adapted to bifurcation; phase jumps are avoided by continuous phase signaling and 
intersymbol interference is mitigated by a cyclic extension; 
(ii) unipolar and bipolar rectified FSK (dubbed RFSK) is proposed and 
investigated w.r.t.~efficiency, output power, and bit error rate; 
(iii) efficiency is studied for the steady state and for symbol transitions; 
symbol transitions potentially have an impact on output power; 
the impact on output power is small as long as the symbol duration is about ten times longer than the effective duration of the impulse response; 
(iv) coupling coefficients causing orthogonal signaling are identified; 
(v) an equivalent discrete-time channel model is formulated for the setup under investigation, and the impulse response 
is computed; 
(vi) analytic results are supported by an experimental prototype; (vii) a low-cost noncoherent receiver is suggested.

The remainder is organized as follows: In Section~\ref{sec2} properties of the voltage transfer function and the impulse response 
of a series-series topology are investigated, given arbitrary parameters.
Based on the impulse response (which is rarely looked at in this context), in Section~\ref{sec3} an equivalent discrete-time 
channel model is formulated, which provides an exact representation of the input/output behavior of the channel including 
pulse shaping and receive filtering.  In Section~\ref{sec4} a low-cost binary 
frequency shift keying system is suggested that is matched to the bifurcation effect.  Hence, data transmission does not significantly 
degrades power transmission, although both are transmitted via the same forward channel. 
A rectified modification of 2-ary FSK, coined RFSK, is proposed in Section~\ref{sec5}. 
It is shown in Section~\ref{sec6} that data transmission based on 2-ary FSK and RFSK causes a negligible loss of output power, 
as long as the symbol duration exceeds the duration of the impulse response by about one order of magnitude. 
Experimental examinations supporting the simulations are shown in Section~\ref{sec7}.  
Finally, conclusions are drawn in Section~\ref{sec8}. 

\section{Magnetic Resonance}\label{sec2}
\begin{figure}[t]
\centering
\subfigure{
\includegraphics[scale=0.44]{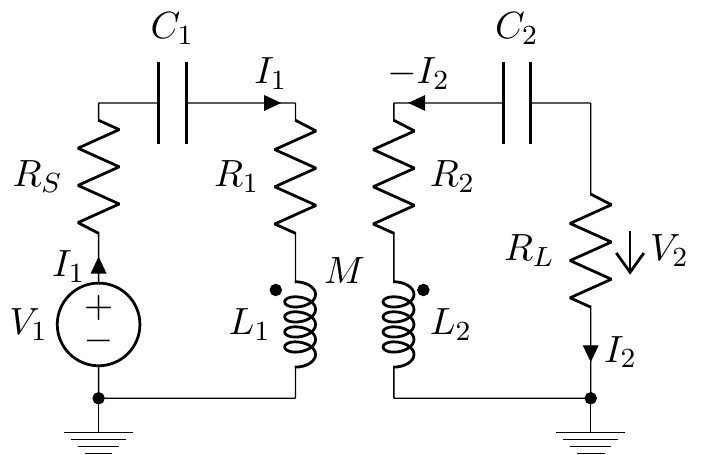}
}
\subfigure{
\includegraphics[scale=0.44]{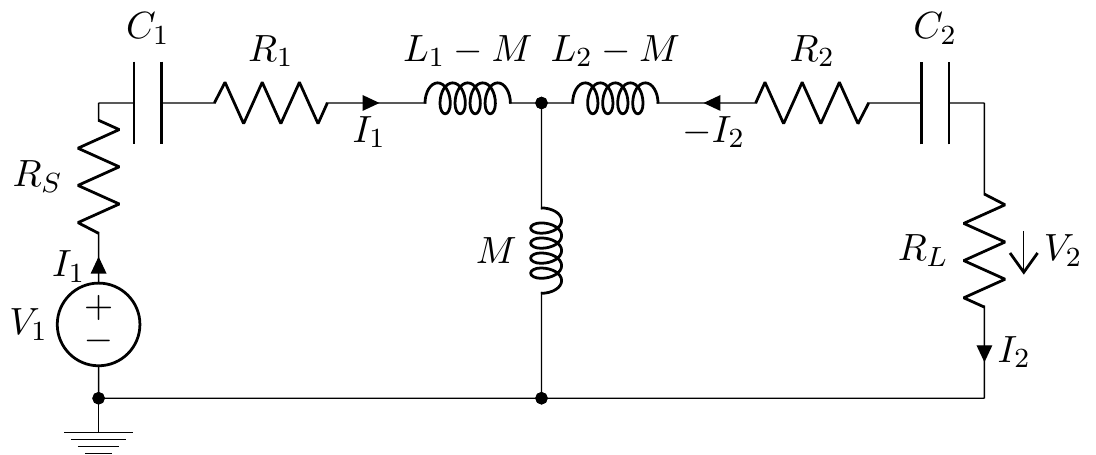}
}
\caption{Two-coil series-resonant circuit model under investigation (left) and equivalent impedance transform (right). 
The equivalent impedance transform is valid if and only if $M>0$.}
\label{fig1}
\end{figure}

For reasons of conciseness, emphasis is on a two-coil resonant system with series-series (SS) topology, c.f.~Fig.~\ref{fig1}. 
Frequency splitting, however, is neither restricted to two-coil systems \cite{zha14b} nor to the SS topology \cite{che13}.
FSK/RFSK is applicable to any circuit type, as long as frequency splitting occurs (but does not collapse otherwise).

For the purpose of analysis, the inner two-port network of the series-resonant circuit model shown on the left side in 
Fig.~\ref{fig1} can be replaced by the T-network structure depicted on the right.  The primary side consists of
capacitance $C_1$ and inductance $L_1$.  Resistance $R_1$ accounts for imperfections.  The same holds for 
the secondary side, marked by index~2. It is desirable that $R_S\ll R_L$, dubbed unsymmetrical system in literature, because otherwise 
an efficient power transfer would not be possible.  The equivalent load resistance, $R_L$, depends on the battery charge condition. 
We assume perfect tuning, i.e., $L_1\,C_1=L_2\,C_2:=L\,C$.  
The mutual inductance is denoted as $M$.  $L_1-M$ and $L_2-M$ are leakage inductances.  The coupling coefficient, $k:=M/\sqrt{L_1\,L_2}$, 
can be interpreted as the normalized mutual inductance, since $0\leq k\leq 1$.  This parameter depends on the geometries 
of the coils, their distance and alignment, and the materials around and in-between the coils.
If $k<k_{\mbox{\footnotesize split}}$, the resonant frequency is $\omega_0=1/\sqrt{LC}$, both for series-resonant and parallel-resonant circuits. 
The quality factor, $Q$, is defined as the ratio of average stored power to dissipated power. $Q$ can be calculated as 
$Q_1=\omega_0 L_1/(R_1+R_S)$ for the primary side and $Q_2=\omega_0 L_2/(R_2+R_L)$ for the secondary side, respectively, see e.g.~\cite{jia15}.  

Different methods have been published in order to define and to determine the peak frequencies $\omega_{-}$ and $\omega_{+}$. 
In most works the splitting behavior analysis is based on circuit theory, but sometimes on an asymptotic coupled-mode theory method \cite{niu12}.
Also, different optimization criteria have been defined.  As mentioned before, {\em frequency splitting} focuses on the output characteristic. 
Usually, the load power is maximized: $\delta P_2/\delta \omega = 0$.  {\em Bifurcation} focuses on the input characteristic. 
Typically, the input impedance $Z_1=V_1/I_1$ seen by the power source is forced to be real-valued: $\mbox{Im}\{Z_1\}=0$.  As a consequence, the phase shift between 
input voltage, $V_1$, and input current, $I_1$, disappears at the peak frequencies.  
This observation has been exploited for 
frequency control \cite{niu13}.  For both optimization criteria, the peak frequencies can be tightly approximated as 
\begin{eqnarray}\label{approx}
\omega_{-} &\approx& \omega_0/\sqrt{1+k},\nonumber\\
\omega_{+} &\approx& \omega_0/\sqrt{1-k},
\end{eqnarray}
see for example \cite[Section~III~F]{niu13}. 
Given these frequencies, the coupling coefficient can be calculated as
\begin{equation}
k\approx \frac{\omega^2_{+}-\omega^2_{-}}{\omega^2_{+}+\omega^2_{-}}.
\end{equation}
Accuracy improves with increasing $k$.  
At $k\leq k_{\mbox{\footnotesize split}}$, $\omega_0=\omega_{-}=\omega_{+}$.
Note that the (possibly time-varying) load $R_L$ has an impact on $k_{\mbox{\footnotesize split}}$, 
but no significant influence on the peak frequencies if $k>k_{\mbox{\footnotesize split}}$. 

In this contribution, focus is on the entire voltage transfer function rather than on the input or output characteristic.
According to Kirchhoff's circuit laws, 
\begin{eqnarray}
V_1(s) &\!\!\!=\!\!\!& R'_S I_1(s) + \frac{I_1(s)}{C_1 s} + L'_1 s I_1(s) + M s (I_1(s)-I_2(s)) \nonumber\\
0 &\!\!\!=\!\!\!& L'_2 s I_2(s) + \frac{I_2(s)}{C_2 s} + R'_L I_2(s) + M s (I_2(s)-I_1(s)) \nonumber\\
V_2(s) &\!\!\!=\!\!\!& R_L I_2(s),
\end{eqnarray}
where $R'_S:=R_S+R_1\approx R_S$, $R'_L:=R_L+R_2\approx R_L$, $L'_1:=L_1-M$, $L'_2:=L_2-M$, and $s:=\sigma+j\omega$ ($\sigma\leq 0$), 
see for example \cite{jia13}.
Hence, the transfer function $H(s) := V_2(s)/V_1(s)$ can be written in normal form as 
\begin{equation}
H(s):=\frac{N(s)}{D(s)}=\frac{a_3\,s^3}{b_4\,s^4+b_3\,s^3+b_2\,s^2+b_1\,s+b_0},
\end{equation}
where $a_3:=R_L M$, $b_4:=L_1 L_2(1-k^2)$, $b_3:=R'_S L_2+R'_L L_1$, $b_2=:R'_S R'_L + L_1/C_2 + L_2/C_1$, $b_1:= R'_L/C_1 + R'_S/C_2$, 
and $b_0:=1/(C_1 C_2)$.
The peak frequencies can be determined exactly by calculating the roots of the denominator polynomial 
\begin{equation}\label{ds}
D(s)=b_4\,s^4+b_3\,s^3+b_2\,s^2+b_1\,s+b_0.
\end{equation}
Two pairs of roots $\sigma_{1/2}\pm j\omega_{1/2}$ exist, $\sigma_1<\sigma_2$.  
If $k\leq k_{\mbox{\footnotesize split}}$, $\omega_1=\omega_2=\omega_0$.  However if $k>k_{\mbox{\footnotesize split}}$, 
$\omega_1=\omega_{+}>\omega_0$ and $\omega_2=\omega_{-}<\omega_0$.

The polynomial factorization simplifies if $D(s)$ is imaginary-valued (since $N(s)\sim s^3$) and therefore $H(s)=N(s)/D(s)$ is real-valued.
The necessary condition is 
\begin{equation}\label{ace} 
b_4\,s^4 + b_2\, s^2 + b_0 = 0
\end{equation}
respectively
\begin{equation}
b_4\,\omega^4 - b_2\, \omega^2 + b_0 = 0.
\end{equation}
The solution is a quadratic function of $\omega$, similar to the one known from bifurcation \cite{niu13}. 
The match between the peak frequencies identified by the roots and by the quadratic function improves with increasing $k$. 

\begin{figure}[t]
\centering
\includegraphics[scale=0.35]{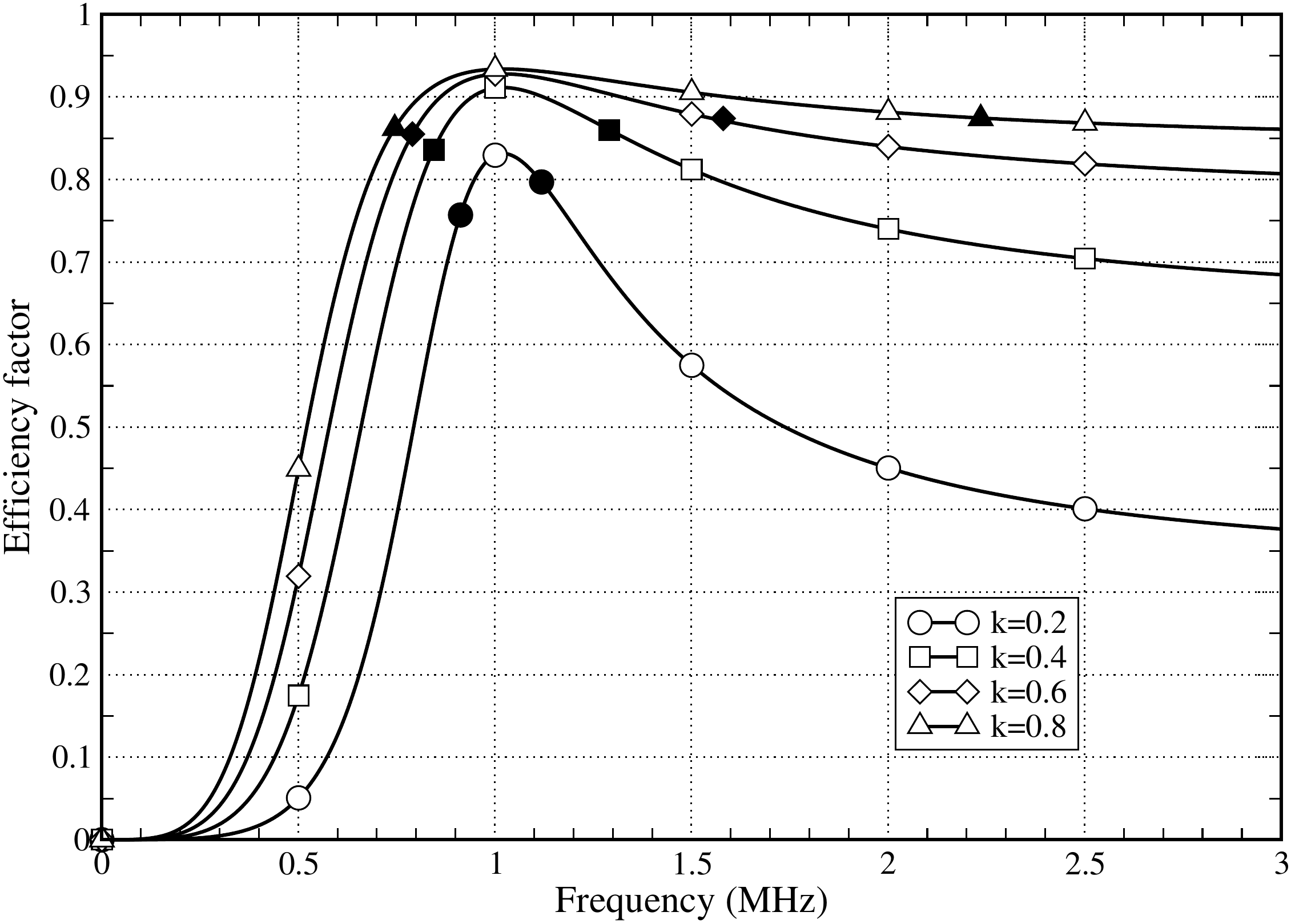}
\caption{Efficiency factor $\eta$ vs.~frequency. Parameters are taken from Fig.~\ref{fplus_fminus_vs_k}. 
Efficiencies at the peak frequencies are marked by filled symbols.}
\label{eta_vs_f}
\end{figure}

However, the goal of this contribution is not to explore ``yet another method'' for calculating peak frequencies.
Substituting (\ref{ace}) into (\ref{ds}) yields 
\begin{eqnarray}
|H(j\omega_{+})| &=& \frac{a_3 \omega_{+}^2}{b_3 \omega_{+}^2-b_1} = \frac{a_3}{b_3}\cdot \frac{\omega_{+}^2}{\omega_{+}^2-b_1/b_3}\nonumber\\
|H(j\omega_{-})| &=& \frac{a_3 \omega_{-}^2}{b_1-b_3 \omega_{-}^2} = \frac{a_3}{b_3}\cdot \frac{\omega_{-}^2}{b_1/b_3-\omega_{-}^2}.  
\end{eqnarray} 
Inserting $b_1/b_3=\omega_0^2$ and (\ref{approx}) into the right-hand side 
proves that the magnitude of the voltage transfer function is identical at both peaks, 
if (\ref{approx}) is fulfilled with equality. 
This observation is useful regarding data transmission: 2-ary FSK is well balanced (if imperfections are neglected).
Note, however, that our analysis is based on an Ohmic load $R_L$.  Non-Ohmic loads cause unbalanced peaks.
The same statement can be drawn concerning output power: Output power $P_2=\mbox{Re}\{V_2\cdot I_2^*\}=|V_2|^2/R_L$ 
is maximized at the peaks of the voltage transfer function. 
Noticeably, the situation is different w.r.t.~the efficiency factor $\eta:=P_2/P_1$, however. 
As shown in Fig.~\ref{eta_vs_f}, efficiency is maximized at the resonant frequency $f_0$ for all coupling coefficients $k$. 
This result has been obtained for the steady state by conducting a network analysis based on Kirchhoff's circuit laws. 
Unlike output power, efficiency does not split. 
Luckily, at the peak frequencies the loss is moderate, unless efficiency is of extreme relevance. 
For the parameters under investigation, $\eta_{-}=0.837$ at $f_{-}$, $\eta_0=0.911$ at $f_0$, and $\eta_{+}=0.861$ at $f_{+}$, respectively. 
The influence of a frequency mismatch on the efficiency is inherently included in this figure. 

\begin{figure*}[t]
\centering
\subfigure{
\includegraphics[scale=0.35]{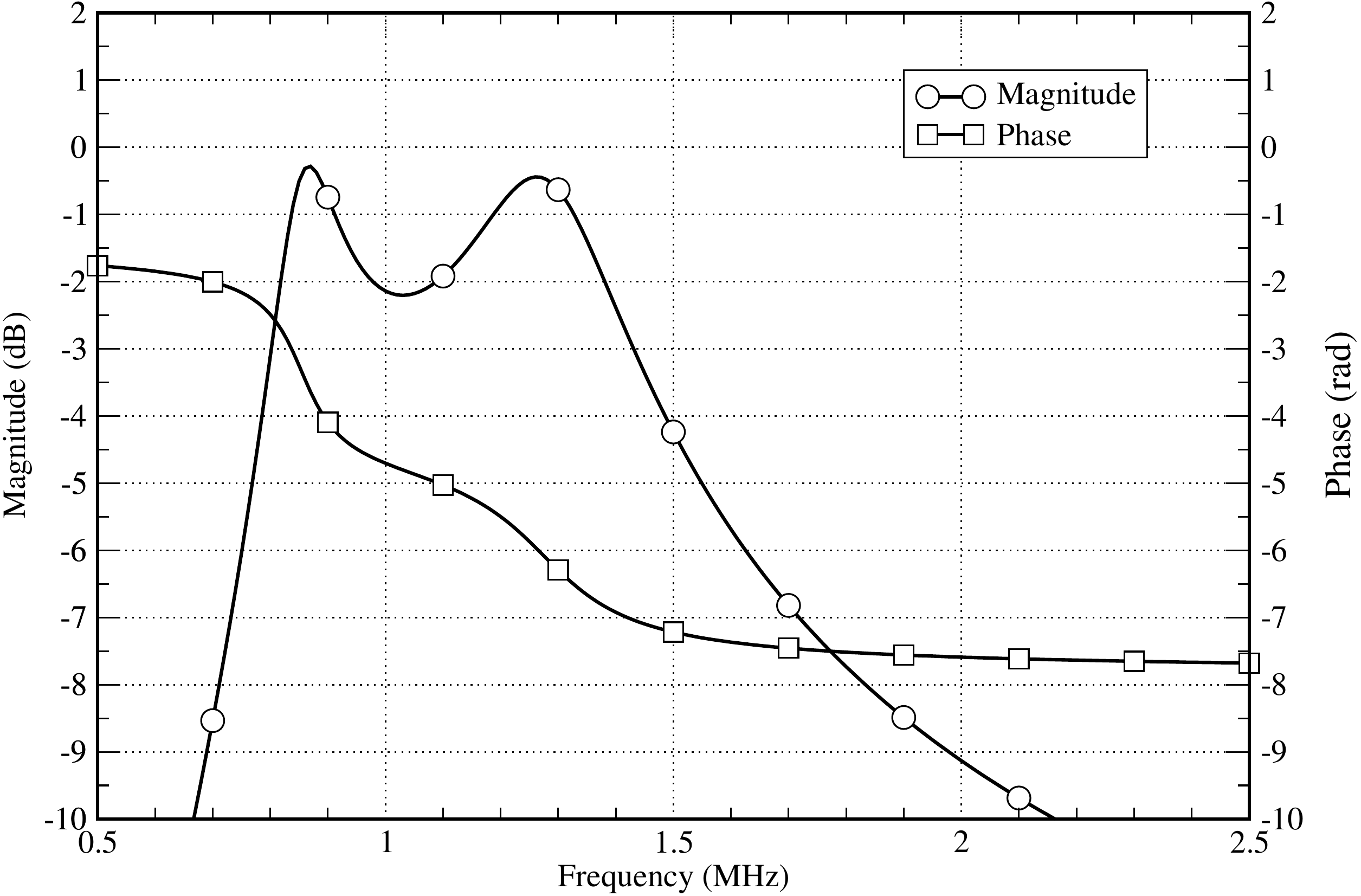}
}~
\subfigure{
\includegraphics[scale=0.35]{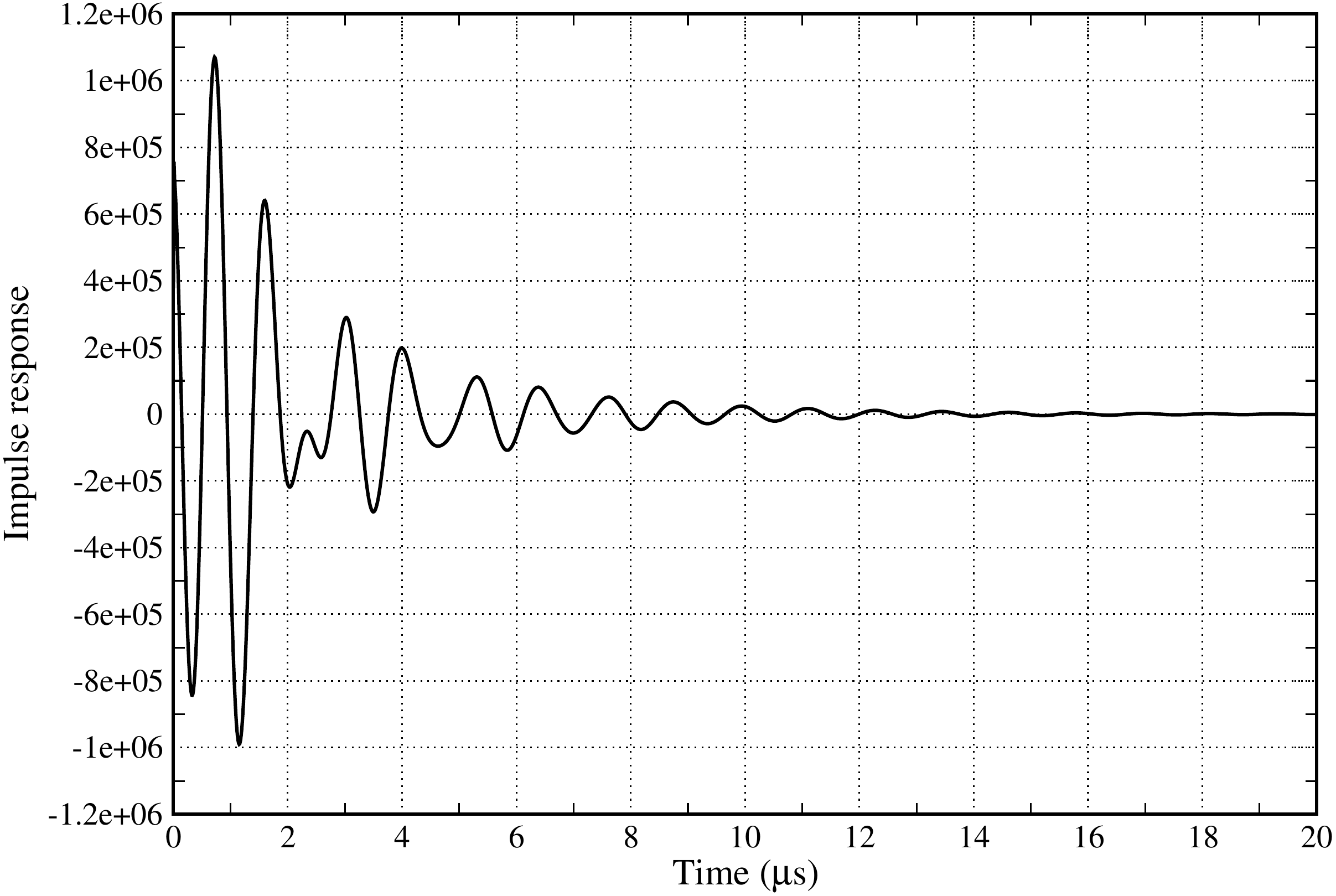}
}
\caption{Voltage transfer function (left) and impulse response (right) for an inductively coupled resonant circuit assuming the following parameters: 
$C_1=C_2=4$~nF, $L_1=L_2=\SI{6.3}{\micro \henry}$, $R_1=R_2=0.62~\Omega$, $R_S=0.17~\Omega$, $R_L=10~\Omega$, $k=0.4$.  The resonant frequency is at 
$f_0=1$~MHz.  The peak frequencies are $f_{-}\approx 0.845$~MHz and $f_{+}\approx 1.291$~MHz, respectively.  Note that the phase of $H(s)$ is $\pi$~rad at $f_{-}$ and 
0~rad at $f_{+}$. 
If the orientation of one coil would be reversed, the phase would shift by $\pi$.}
\label{fig2}
\end{figure*}

On the subject of data communications, the knowledge of the impulse response is important, 
because the impulse response determines the amount of intersymbol interference (ISI).
The causal, real-valued impulse response $h(t)$ is the inverse Laplace transform of the transfer function $H(s)$:
\begin{equation}
h(t) \LEFT H(s).
\end{equation}
In Fig.~\ref{fig2} the transfer function and the corresponding impulse response is featured for an example.
The effective duration of the impulse response, about $5-\SI{10}{\micro s}$ in this example, is determined by the more narrow peak in frequency domain, 
i.e., the peak centered around $f_{-}$. 
Besides characterizing ISI and its impact on data rate and equalization, the impulse response is the key recipe in order to 
derive an equivalent discrete-time channel model (Section~\ref{sec3}).

\section{Equivalent Discrete-Time Channel Model}\label{sec3}
\begin{figure}[t]
\centering
\includegraphics[scale=0.46]{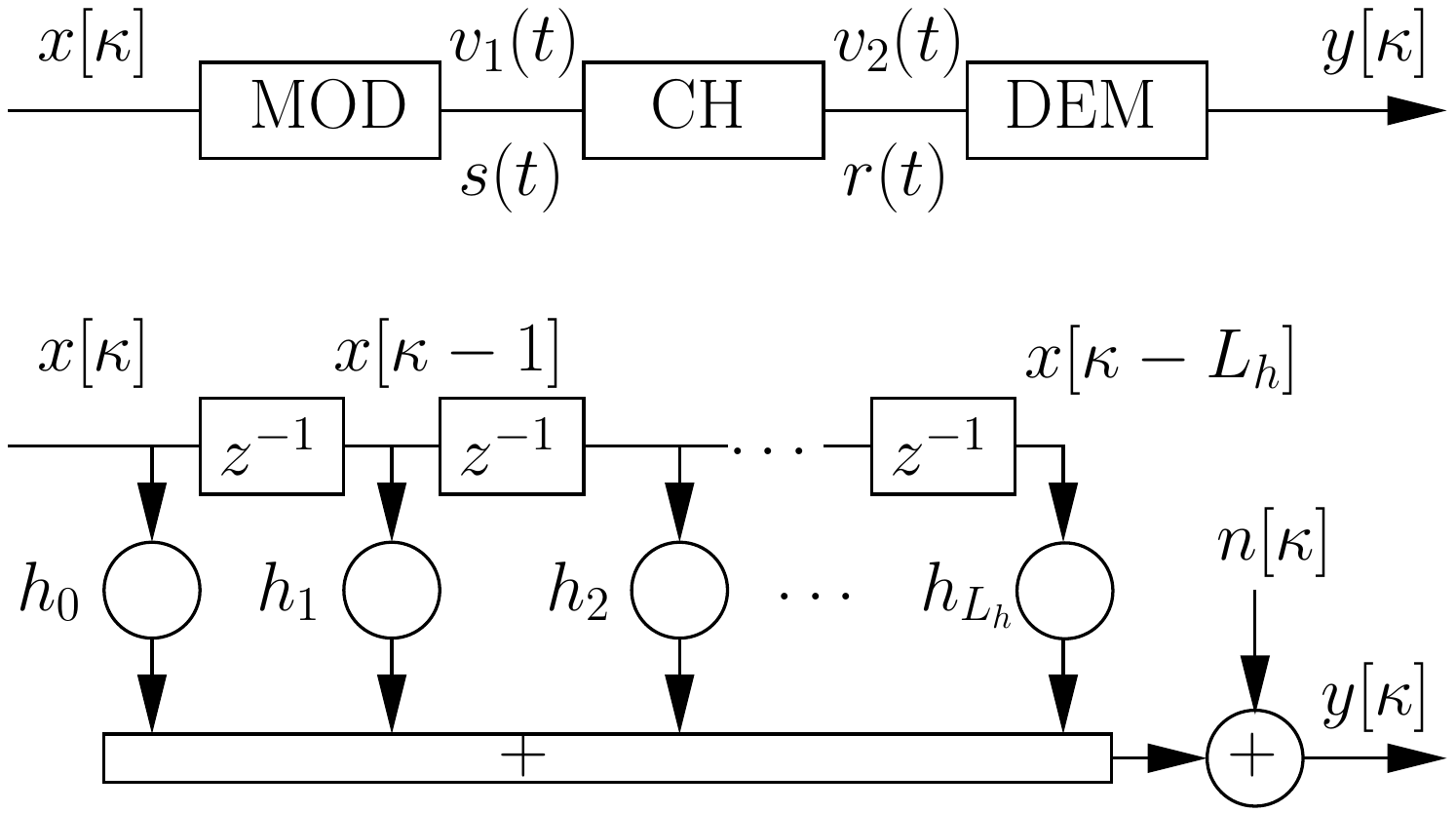}
\caption{Transmission system (top) and equivalent discrete-time channel model (bottom).  Each delay element $z^{-1}$ procures a delay $T_s$.}
\label{fig3}
\end{figure}

In digital communications, it is more convenient to handle discrete-time signals rather than analog signals, particularly in computer simulations 
and for the purpose of analysis. 
The discrete-time channel model shown at the bottom of Fig.~\ref{fig3} represents the input/output behavior of the 
transmission system depicted at the top of the same figure \cite{hoe92}.  
The input/output behavior is exact for all linear transmission systems and channels. 
Inputs are data symbols $x[\kappa]$ (before pulse shaping), 
outputs are samples $y[\kappa]$ (after receive filtering and subsequent sampling at sampling rate $1/T_s:=1/(J\,T)$):
\begin{equation}\label{edtcm}
y[\kappa] = \sum_{l=0}^{L_h} h_l\, x[\kappa-l] + n[\kappa],
\end{equation}
where $\kappa$ is the time index, $T$ the symbol duration, $T_s$ the sample duration, and $J\ge 1$ the oversampling factor.  
The oversampling factor $J$ must be large enough 
in order to fulfill the sampling theorem.  Particularly, the discrete-time channel model must precisely represent the two passbands of 
$H(j\omega)$. We need to distinguish between the desired signal including ISI, and the noise process. 

The ISI is represented by a finite impulse response (FIR) filter, consisting of $L_h+1$ channel coefficients $h_l$ and equally-spaced delay elements.  
The channel coefficients stand for the impulse response.  The FIR filter is fed with data symbols.  Two neighboring data symbols 
$x[\kappa]$ and $x[\kappa+J]$ are padded by $J-1$ zeros.  The channel coefficients $h_l$, $0\leq l\leq L_h$, are causal, real-valued and time-invariant. 
The latter is due to the fact that in the near field no multipath fading takes place, because the wavelength exceeds all 
relevant distances.  Hence, there is no destructive superposition of waves.  The effective memory length $L_h$ is defined so that additional coefficients 
do not contribute to the performance.  The channel coefficients can be derived as follows. Given the transfer function in analog domain, $H(s)$, 
the channel coefficients $h_l$ can be derived exactly by means of the bilinear transformation as
\begin{equation}
H(z) = H(s)\big|_{s=\frac{2}{T_s} \frac{z-1}{z+1}}
\end{equation}
followed by the inverse $z$-transform
\begin{equation}
h_l\big|_{0\leq l\leq L_h} \LEFT H(z).
\end{equation}
Alternatively, the inverse Laplace transform 
\begin{equation}
h(t) = \frac{1}{2\pi\,j}\, \int_{-j\infty}^{j\infty} e^{st}\, H(s)\, ds
\end{equation}
can be analytically or numerically evaluated at samples $t=l\,T_s$, $l\ge 0$. 
Another alternative is to perform an inverse discrete Fourier transform of the sampled spectrum $H(j\omega)$. 
The effective length of the impulse response is a function of the quality factor $Q$ \cite{kis15}. 

The noise statistics depends on the origin of the noise.  
We distinguish between the noise generated at the primary side, $n_1(t)$, and the noise affecting the secondary side, $n_2(t)$, 
including interference picked up from the magnetic field.  The latter contribution may be even dominant in harsh environments.
The noise process generated by the power source is often overlooked in literature.  It refers to the difference between the true input signal $v_1(t)$ 
and the desired input signal (e.g., an ideal sine wave).  Rectification causes nonlinear distortions. 
The noise process generated by the load $R_L$ is often assumed to be caused by thermal noise, also known as Johnson noise. 
The thermal noise components produced by $R_S$, $R_1$ and $R_2$ are typically small, because Johnson noise is proportional to the resistance. 
The common noise process can be modeled as
\begin{equation}
n(t) = \sigma_1\, h(t)*n_1(t) + \sigma_2\, n_2(t),
\end{equation}
where $\lim\limits_{T\to\infty} \frac{1}{T} \int_0^T h^2(t)\,dt=1$, $\lim\limits_{T\to\infty} \frac{1}{T} \int_0^T n_1^2(t)\,dt=1$, 
and $\lim\limits_{T\to\infty} \frac{1}{T} \int_0^T n_2^2(t)\,dt=1$ without loss of generality.
The noise powers (and hence the signal-to-noise ratio) can be controlled by the variances $\sigma_1^2$ and $\sigma_2^2$. 
In the special case of white noise at the primary side, the power spectral density of this noise component is proportional to $|H(j\omega)|^2$.
This case simplifies analysis and optimum transmitter design, because it would result in a constant spectral signal-to-noise ratio 
(since signal power and noise power are proportional for all frequencies). 
In the case of white noise at the secondary side, analysis simplifies as well.

\section{FSK System Design Matched to Frequency Splitting}\label{sec4}
Throughout this article focus is on the forward channel, i.e., 
on the link from the primary side to the secondary side, 
because attention is on modulating data with little impact on power transmission.  Sometimes, it is also important for wireless 
power transmission systems to transmit information in reverse direction as well.  Separation of forward link and feedback link is
a classical duplexing problem, which is beyond the scope of this paper.  Time-division duplexing (based on orthogonal time slots) 
is one of several possibilities.  Frequency-division duplexing (employing different frequency bands) is another option.  But also techniques making use of mutual coupling, like load-shift keying, are common.

Given a dispersive channel distorted by additive white Gaussian noise, w.r.t.~data transmission 
orthogonal frequency-division multiplexing (OFDM) with water-filling is capacity achieving  
\cite{won99, kis15}.  Simply speaking, this is due to the fact that 
the power of each OFDM subcarrier can be matched individually to the voltage transfer function $H(j\omega)$.  Since the inductive channel is time-invariant 
in the near field when transmitter and receiver are stationary, water-filling and bit-loading are relatively easy to implement.  
Also, a cyclic extension can easily be realized 
in conjunction with OFDM in order to transform the dispersive channel into parallel non-dispersive subchannels \cite{nee00}.  
The cyclic extension causes some loss of power efficiency (w.r.t.~data detection, not w.r.t.~energy harvesting) 
and some loss of bandwidth efficiency, but for a large number of subcarriers this loss is negligible.  
In time-invariant environments the number of subcarriers is not limited by the symbol duration. 
In the given application, the complex-valued OFDM signal must be transformed into a real-valued signal 
by means of a quadrature modulator or a similar device. 
Alternatively, baseband processing can be performed when using a real-valued multicarrier variant like discrete multitone transmission (DMT). 
Then the bandwidth efficiency of data transfer is reduced by a factor of two.  The main problem in this context, 
however, is the high peak-to-average power ratio (PAPR), making OFDM/DMT inefficient w.r.t.~power transfer. 
Although a vast number of techniques have been published in order to reduce the PAPR \cite{jia08}, neither OFDM nor DMT have constant amplitude. 
Hence, we must select an alternative modulation scheme when the impact of data transfer on a simultaneous power transfer should be small. 

Classically, there is a trade-off concerning the quality factor $Q$: A high $Q$ value is desired for efficient power transfer, 
whereas a low $Q$ value is superior for data transfer, because a low $Q$ value means a larger bandwidth (at the cost of distance). 
In this contribution, perhaps for the first time we opt for a high $Q$ value both for data and power transfer assuming that 
$k>k_{\mbox{\footnotesize split}}$. 
Given a fixed distance (and hence constant coupling coefficient $k$), frequency splitting improves with increasing $Q$. 
For a high $Q$ value, the two peaks are containing most of the energy.  This fact motivates a 2-ary FSK system design: 
Data bit $u[\kappa]=0$ is mapped onto $f_{-}$ (called ``space''), whereas data bit $u[\kappa]=1$ is mapped onto $f_{+}$ (called ``mark'').  
Hence, the power-normalized transmit signal is of the form 
\begin{equation}\label{s}
s(t) = \left\{ \begin{array}{cl}
\sqrt{2}\, \sin(2\pi f_{-}t+\phi) \quad & \mbox{if}~u[\kappa]=0 \\
\sqrt{2}\, \sin(2\pi f_{+}t+\phi) \quad & \mbox{if}~u[\kappa]=1, \end{array} \right.
\end{equation}
where the cyclic extension of the first symbol is defined over the interval $[0, T_g)$ and the useful symbol duration $T_u$ over $[T_g,T_g+T_u)$.  
The design parameter $T_g$ is the length of the cyclic extension.  If $T_g$ exceeds the effective length of the impulse 
response, ISI does not degrade system performance if receiver-side detection is performed within the range $[T_g, T_g+T_u)$, 
i.e., for the useful symbol duration.  
The symbol rate is $1/T:=1/(T_u+T_g)$.  The cyclic extension does not has a negative impact on the power transfer efficiency 
since the transmit power is constant.  It is a low-effort alternative to equalization. 
The arbitrary phase $\phi$ should be optimized for each symbol so that phase jumps between neighboring symbols are avoided, 
known as continuous phase frequency shift keying (CPFSK). 
Even more smooth transitions can be obtained by employing partial-response continuous phase modulation (CPM) \cite{cpm86}, 
which is beyond the scope of this contribution, however. 
In order to fit to the equivalent discrete-time channel model, $s(t)$ can be written as
\begin{equation}
s(t) = \sqrt{2}\, \sin(2\pi (\overline{f_0}+x[\kappa]\, \Delta f) t+\phi),
\end{equation}
where $\overline{f_0}=(f_{+}+f_{-})/2$ is the effective center frequency, $\Delta f=(f_{+}-f_{-})/2$ the frequency swing, 
and \mbox{$x[\kappa]=2u[\kappa]-1\in \{+1, -1\}$} the binary channel input symbol. 
Note that 2-ary FSK has a constant amplitude (which is maintained on the channel when both peaks have the same gain).
FSK can be interpreted as a special case of OFDM/DMT with two subcarriers. 
In this work we do not tackle the case that $k\leq k_{\mbox{\footnotesize split}}$. 
$M$-ary phase shift keying (PSK) in conjunction with rectangular pulse shaping 
is a constant-amplitude candidate for that case. 
One may switch between FSK, PSK, and perhaps other modulation schemes. 
As pointed out in the numerical results, however, FSK does not collapse if $k<k_{\mbox{\footnotesize split}}$ or if the peak amplitudes are unbalanced.

In the SWIPT literature, the two most frequently discussed resource sharing techniques are called time switching and power splitting, 
respectively \cite{liu13,kri14}.  According to the former strategy, power and data is transmitted in separate time slots.  In the latter strategy, 
the received power is splitted in energy harvesting and data detection.  In this contribution, 
we opt for power splitting.  Contrary to energy harvesting in the $\mu$W regime \cite{liu13,kri14}, however, 
only a very small fraction of the received power 
is needed for data detection.  By means of a high-impedance operational amplifier attached to the secondary coil, 
the data detection unit can be decoupled almost perfectly from the energy harvesting unit. 

Performance of 2-ary FSK is best for coherent reception and orthogonal signaling \cite{cpm86}.  Given the received signal 
\begin{equation}
r(t) = [s(t)+n_1(t)] * h(t) + n_2(t)
\end{equation}
and a sufficiently long cyclic extension, a coherent receiver is of the form 
\begin{equation}\label{mf} 
y[\kappa] = \frac{1}{T_u} \int\limits_{t=T_g+\kappa T}^{(\kappa+1)T} r(t)\cdot \sin(2\pi (\overline{f_0}+\tilde x[\kappa]\, \Delta f) t+\hat \phi)~dt 
\end{equation}
assuming hypotheses $\tilde x[\kappa]\in \{+1, -1\}$, c.f.~Fig.~\ref{fig3}.
In the absence of a cyclic extension, the bit error rate (BER) for binary orthogonal signaling observed in additive white Gaussian noise is \cite{cpm86}
\begin{equation}
P_b = \frac{1}{2} \mbox{erfc}\, \sqrt{\frac{E_s}{2N_0}},
\end{equation}
where $E_s$ is the energy per data symbol and $N_0$ the single-sided noise power density.  
The BER is 3~dB worse than for binary antipodal signaling.  The power loss (w.r.t.~data detection) due to the cyclic extension is 
$10\log_{10}(T/T_u)$~dB.  It is interesting to note that binary orthogonal signaling is accomplished when $f_{+}$ is an integer multiple of $f_{-}$. 
Assuming that the tight approximation (\ref{approx}) is fulfilled with equality, 
\begin{equation}
\frac{f_{+}}{f_{-}} = \frac{\sqrt{1+k}}{\sqrt{1-k}}
\qquad \Leftrightarrow \qquad 
k = \frac{\left(\frac{f_{+}}{f_{-}}\right)^2-1}{\left(\frac{f_{+}}{f_{-}}\right)^2+1}.
\end{equation}
Therefore, orthogonal signaling is reached for the following coupling coefficients:
\begin{equation}
k_{\mbox{\footnotesize orth}} = 3/5, ~8/10, ~15/17, ~24/26, ~35/37, ~48/50, ~63/65, \dots
\end{equation}
These special coupling coefficients hold for arbitrary system parameters. 
Note that at $k=3/5=0.6$ we get $f_{+}/f_{-}=2$ and at $k=8/10=0.8$ we obtain $f_{+}/f_{-}=3$, c.f.~Fig.~\ref{fplus_fminus_vs_k}.
Luckily, BER degradation is small for any other coupling coefficient $k$ as long as $k>k_{\mbox{\footnotesize split}}$, 
as shown in the numerical results in Section~\ref{sec6}.

\section{Rectified Frequency Shift Keying (RFSK)}\label{sec5}
Binary FSK is a fairly simple modulation scheme, but the waveform is fluctuating at any time. 
This has a negative impact on transmitter-side complexity and on hardware efficiency.
In low-power applications, usually a linear power amplifier and a digital-to-analog converter (DAC) are applied.
A linear power amplifier naturally has a worse efficiency factor than a switch. 
Also the DAC accounts to the overall power loss.
These transmitter-side bottlenecks can be avoided by rectifying the sinusoidal transmit signal before feeding it to the resonant circuit. 
The normalized rectified transmit signal, 
\begin{equation}\label{s_rect}
s_{\mbox{\footnotesize rect}}(t) := \mbox{sgn}(s(t)) = \left\{ \begin{array}{cl}
+1 \quad & \mbox{if}~s(t)\ge 0 \\
-1 \quad & \mbox{else}, \end{array} \right.
\end{equation}
has a peak-to-average power ratio of one.  
Bipolar RFSK can be implemented by a full-bridge converter. 
Hence, the maximum possible source power is delivered to the circuit input at low hardware cost. 
Note that the optional cyclic extension is included in (\ref{s_rect}), as well as the design parameter $\phi$ to achieve phase-continuous modulation. 
Power efficiencies $(\eta_{-}+\eta_{+})/2$ 
of 2-ary FSK and bipolar RFSK are about the same for situations with high $Q$, because the harmonics caused by the RFSK waveform 
are sufficiently suppressed by the resonant circuit and therefore have a minor impact on the overall power budget.
If hardware losses of the power converter would be taken into account, RFSK is expected to be more efficient than FSK, 
since a modulated square wave can be generated more efficiently in practice compared to a modulated sinusoidal waveform.
Furthermore, bipolar RFSK generates a higher output voltage 
(and hence output power) given the same peak input voltage.  Consequently, SNR is higher for a peak input voltage constraint. 

Upon zero-level clipping of bipolar RFSK, unipolar RFSK is obtained. 
Complexity is even less, because a half-bridge converter is sufficient. 
Power efficiency is the same as for bipolar RFSK, but output power is one-quarter. 

Although square-wave signals are common in power electronics \cite{kaz95, bur04} and albeit a bipolar rectangular FSK signal is created in \cite{cho13} 
at the receiver side for easing synchronization, RFSK potentially is a novel modulation scheme, at least in conjunction with 
design parameter $\phi$ to achieve phase-continuous modulation, in conjunction with a cyclic extension, and adaptive matching to the bifurcation effect.
According to Fourier analysis, $s_{\mbox{\footnotesize rect}}(t)$ consists of spectral lines at $f_{-}$ and $f_{+}$, respectively, 
and odd multiples thereof. 
Given sufficiently large coupling and quality factors $k$ and $Q$, $H(j\omega)$ is a bandpass filter with two narrow passbands. 
Consequently, the harmonics caused by the RFSK waveform are significantly suppressed by the bandpass, without needing any additional circuit devices.
Assuming the parameters defined in Fig.~\ref{fplus_fminus_vs_k}, the analytical efficiency plotted in Fig.~\ref{eta_vs_f} is obtained.
In Section~\ref{sec7}, the practical efficiency (including hardware imperfections that are not considered in the analysis) is measured for the prototype under investigation.
The temporal efficiency during symbol transitions and the impact of rectification of the sinusoidal transmit signal on the BER will be studied next. 

\section{Numerical Results for FSK and RFSK}\label{sec6}
\begin{figure}[t]
\centering
\subfigure{
\includegraphics[scale=0.35]{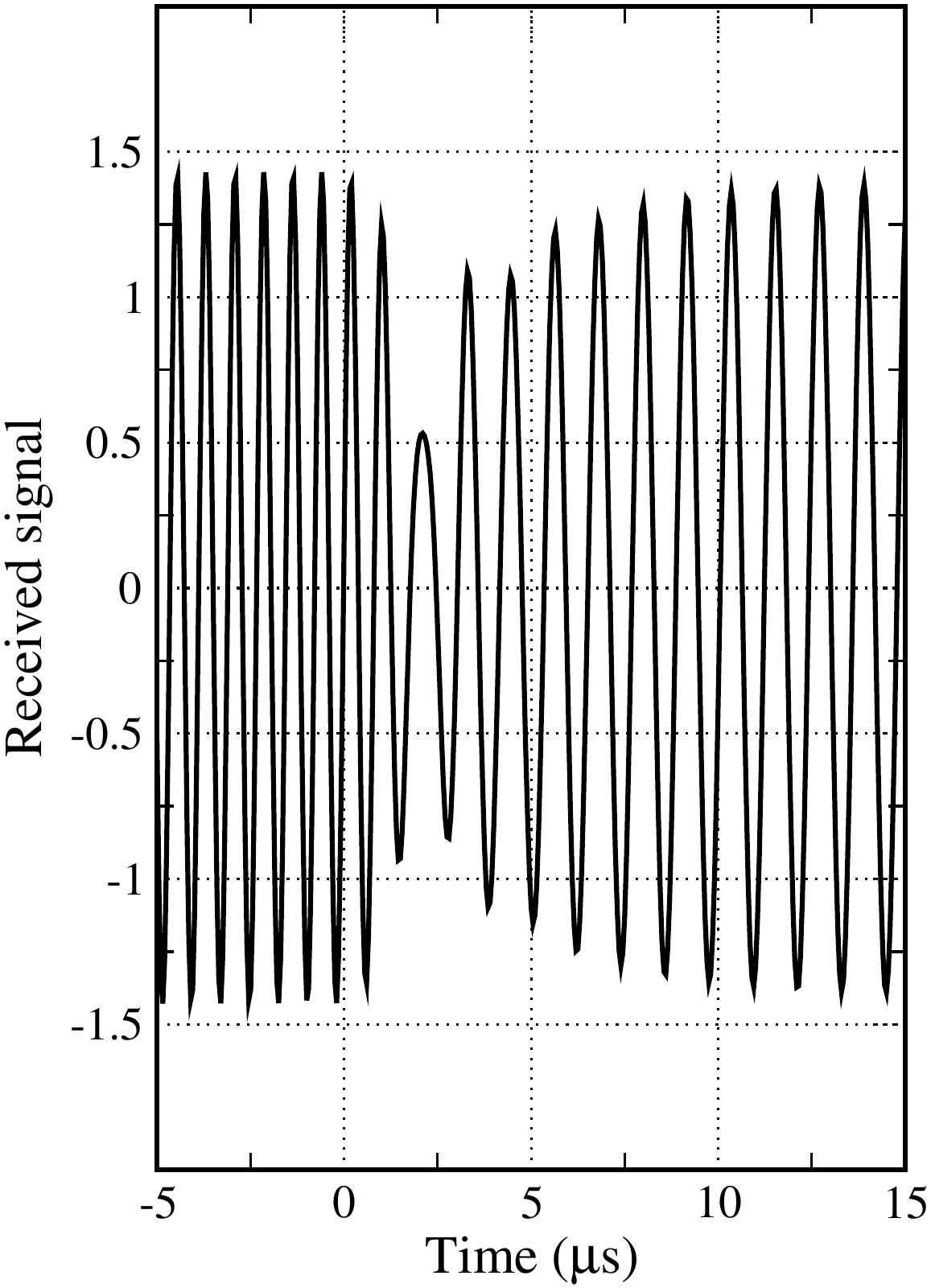}
}~
\subfigure{
\includegraphics[scale=0.35]{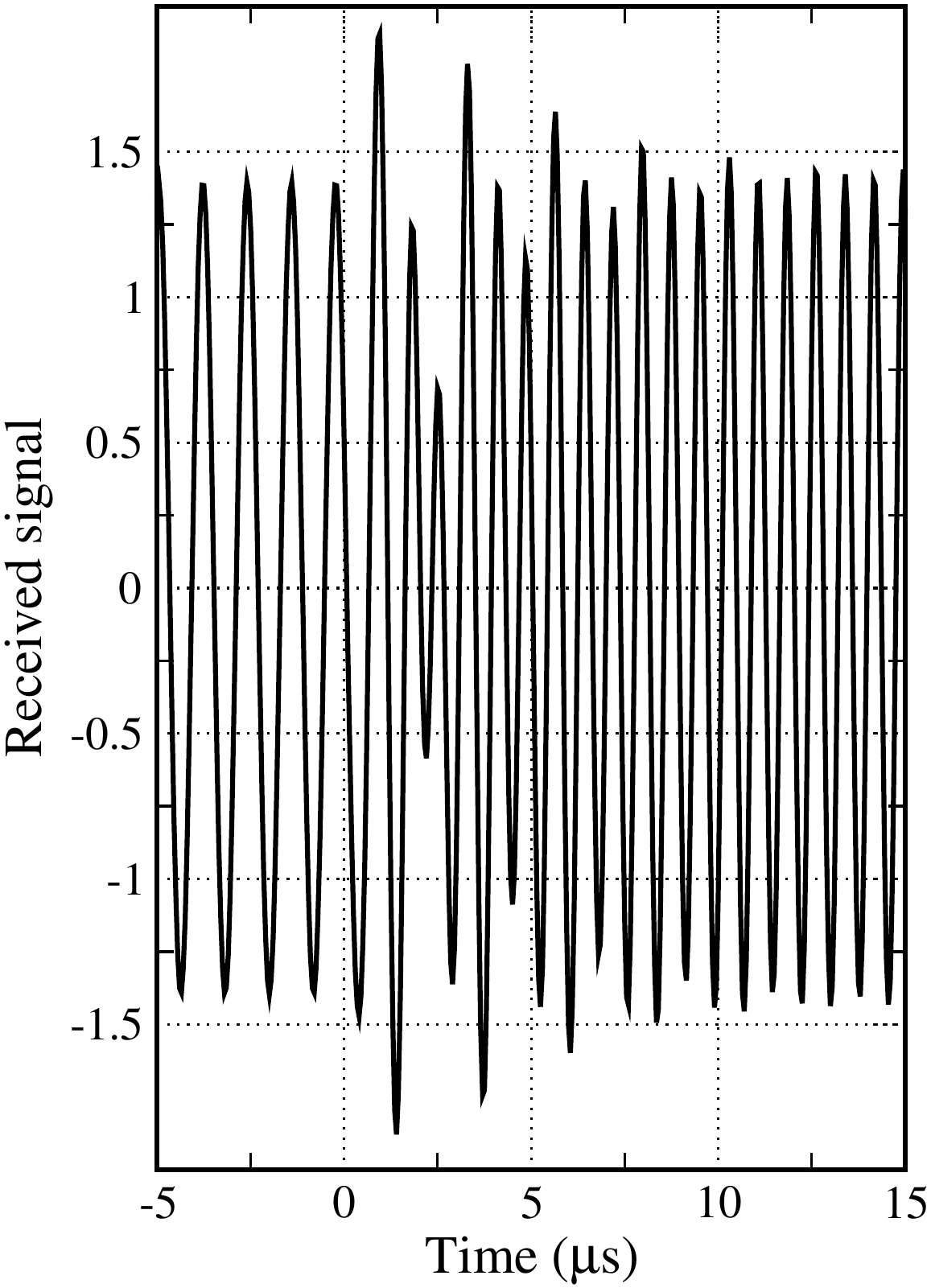}
}
\caption{Noiseless normalized received signal $v_2(t)$ in time domain for 2-ary FSK modulation.  
Left: Transition from $f_{+}$ to $f_{-}$.  Right: Transition from $f_{-}$ to $f_{+}$.
Transition starts at $t=0$ and takes about $5-\SI{10}{\micro s}$.  
Parameters are taken from Fig.~\ref{fig2}.}
\label{signal}
\end{figure}

In the case of 2-ary FSK and RFSK modulation with proper frequency control, 
the received signal is sinusoidal with constant amplitude, except for a transition period when switching from $f_{+}$ to $f_{-}$ or $f_{-}$ to $f_{+}$, 
respectively.  The duration of the transition period is identical with the effective length of the impulse response, c.f.~Fig.~\ref{signal}. 
The output power is not significantly degraded if the symbol duration $T$ is at least ten times the effective length 
of the impulse response.  But how about temporal efficiency just after toggling data symbols?  
In order to investigate this question, we first define the efficiency factor for the transient phase as 
\begin{equation}\label{eta_T}
\eta_T := \frac{\int_0^T v_2(t)\cdot i_2(t)~dt}{\int_0^T v_1(t)\cdot i_1(t)~dt}.
\end{equation}
Note that $\lim\limits_{T\to\infty} \eta_T\to\eta$, where $\eta=P_2/P_1$ according to Section~\ref{sec2}.
The four waveforms in (\ref{eta_T}) have been emulated with a SPICE electronic circuit simulator for 
$T=\SI{10}{\micro s}$, which corresponds to 100~kbps.  Sufficient oversampling is provided.
For the transition $f_{+}\to f_{-}$ we obtain $\eta_T=0.734$.  As expected, this value is smaller than $\eta_{-}=0.837$ at $f_{-}$.
Interestingly, for the transition $f_{-}\to f_{+}$ we get $\eta_T=0.964$ (!), which is larger than $\eta_{+}=0.861$ at $f_{+}$.  
The average efficiency of the transient phase, $\overline{\eta_T}=0.849$, is the same as the average efficiency of the steady state, 
$(\eta_{-}+\eta_{+})/2$.  
The same conclusion has been verified for $T=\SI{100}{\micro s}$ (10~kbps) and $T=\SI{1000}{\micro s}$ (1~kbps), respectively.
This nice result can be explained by the fact that in the transient phase energy oscillates between the circuit components. 
All numbers reported here hold for sinusoidal as well as square-wave signal generators. 

\begin{figure*}[t]
\centering
\subfigure{
\includegraphics[scale=0.35]{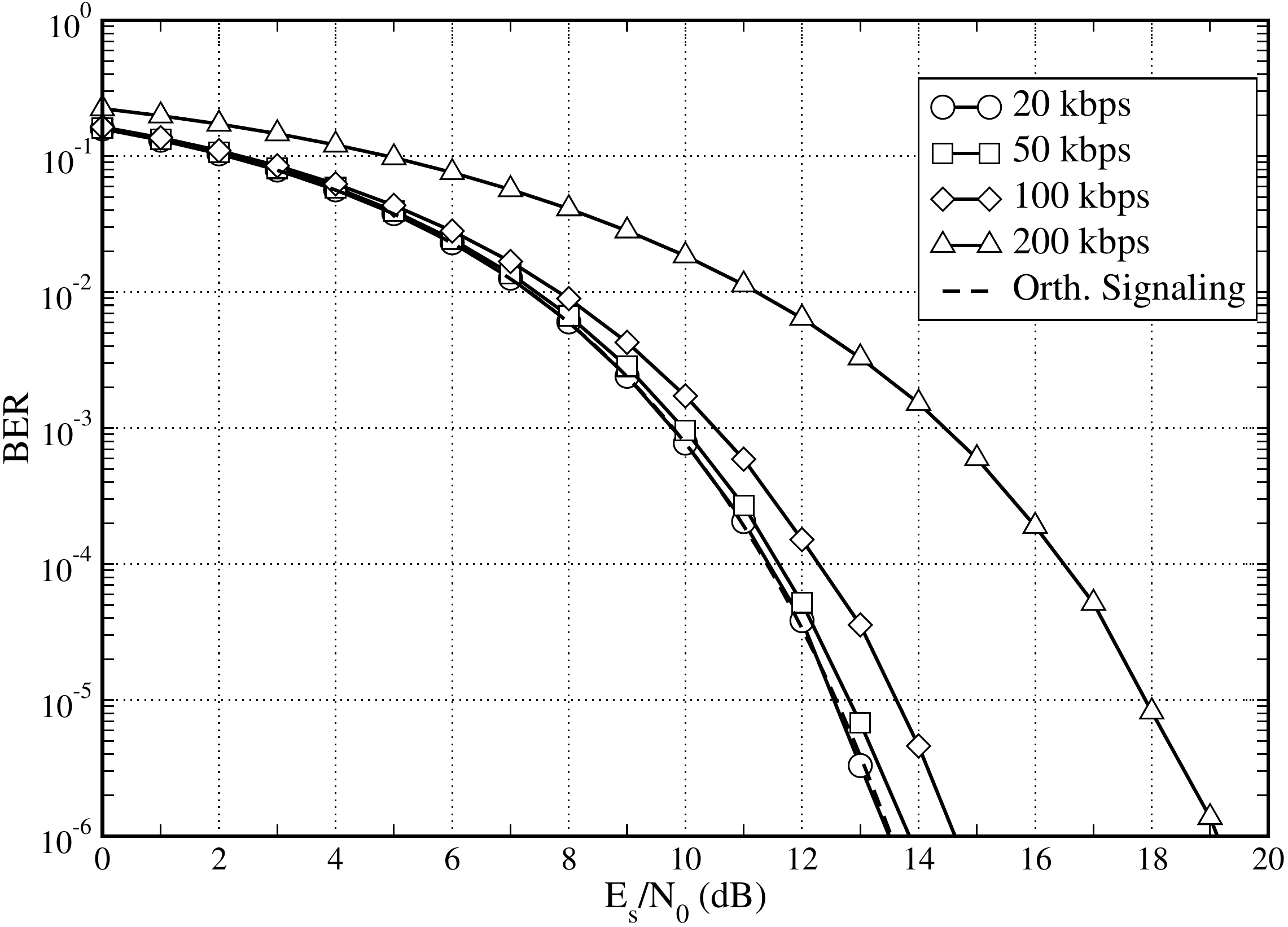}
}\qquad
\subfigure{
\includegraphics[scale=0.35]{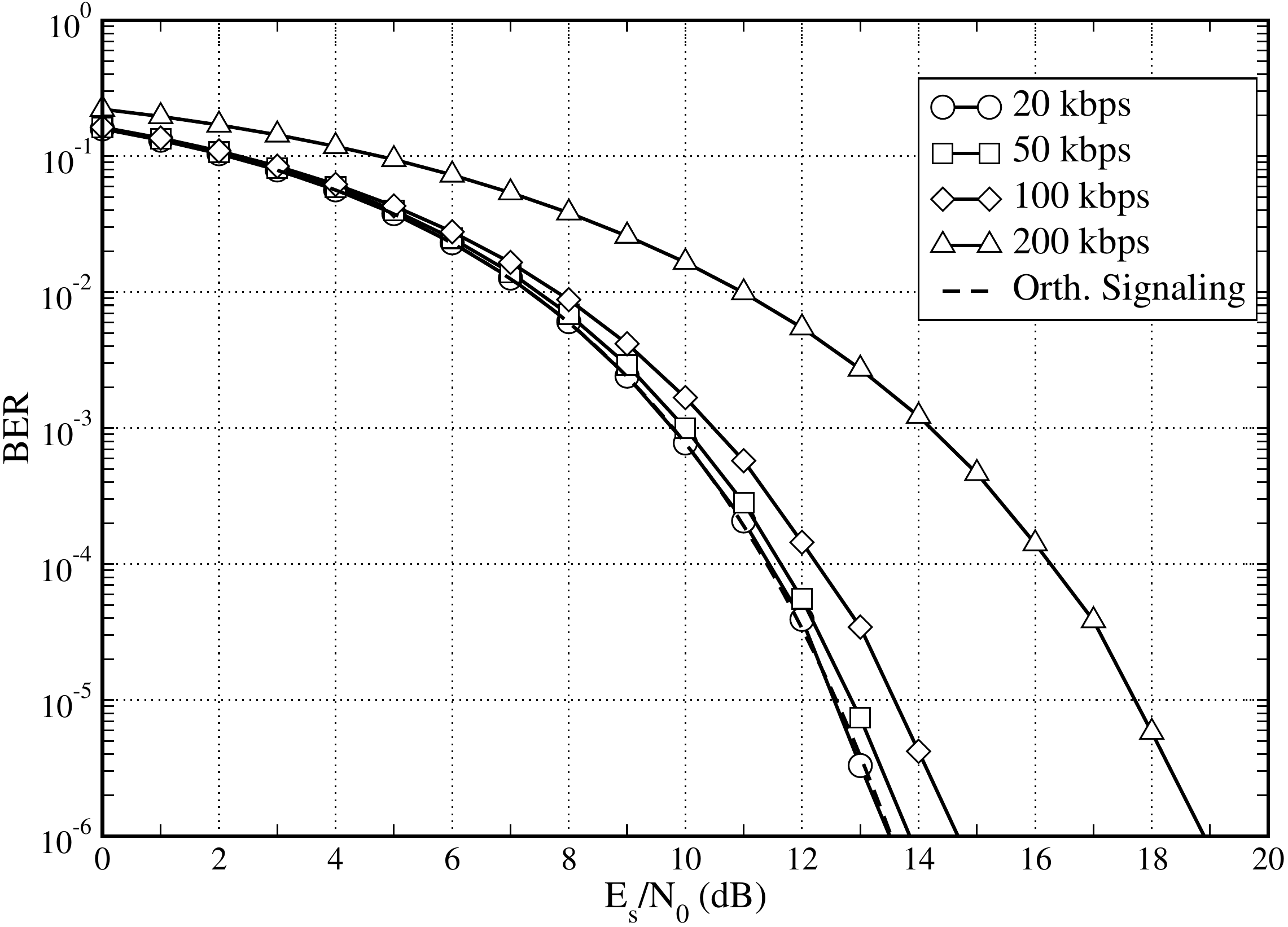}
}
\caption{Bit error rate vs.~SNR for 2-ary FSK modulation (left-hand side) and for rectified FSK (right-hand side) given data rates between 20~kbps and 200~kbps.
The performance of binary orthogonal signaling is depicted by the dashed line.  In all cases an average power constraint is assumed. 
Parameters are taken from Fig.~\ref{fig2}.}
\label{ber}
\end{figure*}

Monte Carlo BER results have been obtained for a coherent receiver exploiting a cyclic extension.  
Phase and clock synchronization are assumed to be perfect, 
as any state-of-the-art synchronization scheme can be applied in this non-fading application. 
In Fig.~\ref{ber}, the BER for 2-ary FSK and bipolar RFSK is shown for data rates between 20~kbps and 200~kbps, 
given the parameters introduced in 
Fig.~\ref{fig2}.  The cyclic extension is chosen to be of fixed length $T_u/10$.  Correspondingly, at 20~kbps the cyclic extension is of length $\SI{5}{\micro s}$, 
which is of the same order as the effective length of the impulse response, c.f.~Fig.~\ref{fig2}.  
Hence, ISI is avoided almost completely in this example. 
For classical 2-ary FSK, the BER is close to binary orthogonal signaling, although $k=0.4 \not= k_{\mbox{\footnotesize orth}}$. 
At 100~kbps (200~kbps), the cyclic extension is just $\SI{1}{\micro s}$ ($\SI{0.5}{\micro s}$) long.  Still, no error floor is visible in the interesting range of bit error rates.
The BER is identical whether the noise process is added at the transmitter or at the receiver side, respectively, 
as long as the noise statistics is the same.  It is noteworthy to mention that we are not aiming high data rates in this contribution.
Data rates can be enhanced by equalization. 
At 200~kbps, the data-rate-to-carrier-frequency ratio is 20~\%. 
For comparison, in \cite{gho04} a data-rate-to-carrier-frequency ratio of up to 67~\% has been reported for 2-ary FSK. 
However, in \cite{gho04} target is at low quality factor $Q$, at the expense of a low power transmission efficiency. 

The fact that binary FSK and RFSK are well matched to bifurcation does not mean that these schemes necessarily fail in situations 
where; (i) the peaks are not well balanced or when (ii) $k<k_{\mbox{\footnotesize split}}$.  As an example of case (i) we have tested 
$k=0.2$ (instead of $k=0.4$).  In this situation $|H(j\omega_{-})|>|H(j\omega_{+})|$.  As an example of case (ii) we have tested $R_L=40~\Omega$ 
(instead of $R_L=10~\Omega$).  In these cases, data rates of 300~kbps can be achieved at $E_s/N_0\approx 19$~dB and $E_s/N_0\approx 17$~dB, respectively, 
both at $10^{-6}$.  Binary FSK and RFSK are quite rugged against imperfections.

Concerning a fair comparison of conventional 2-ary FSK and rectified FSK, 
we must distinguish between an average power constraint and a peak input voltage constraint.  
For an average power constraint, in our example the BER is not distinguishable for both versions, see Fig.~\ref{ber}.
For a peak input voltage constraint, however, bipolar RFSK would benefit from a gain, because of the factor $\sqrt{2}$ in (\ref{s}). 

\begin{figure*}[t]
\centering
\subfigure{
\includegraphics[scale=0.35]{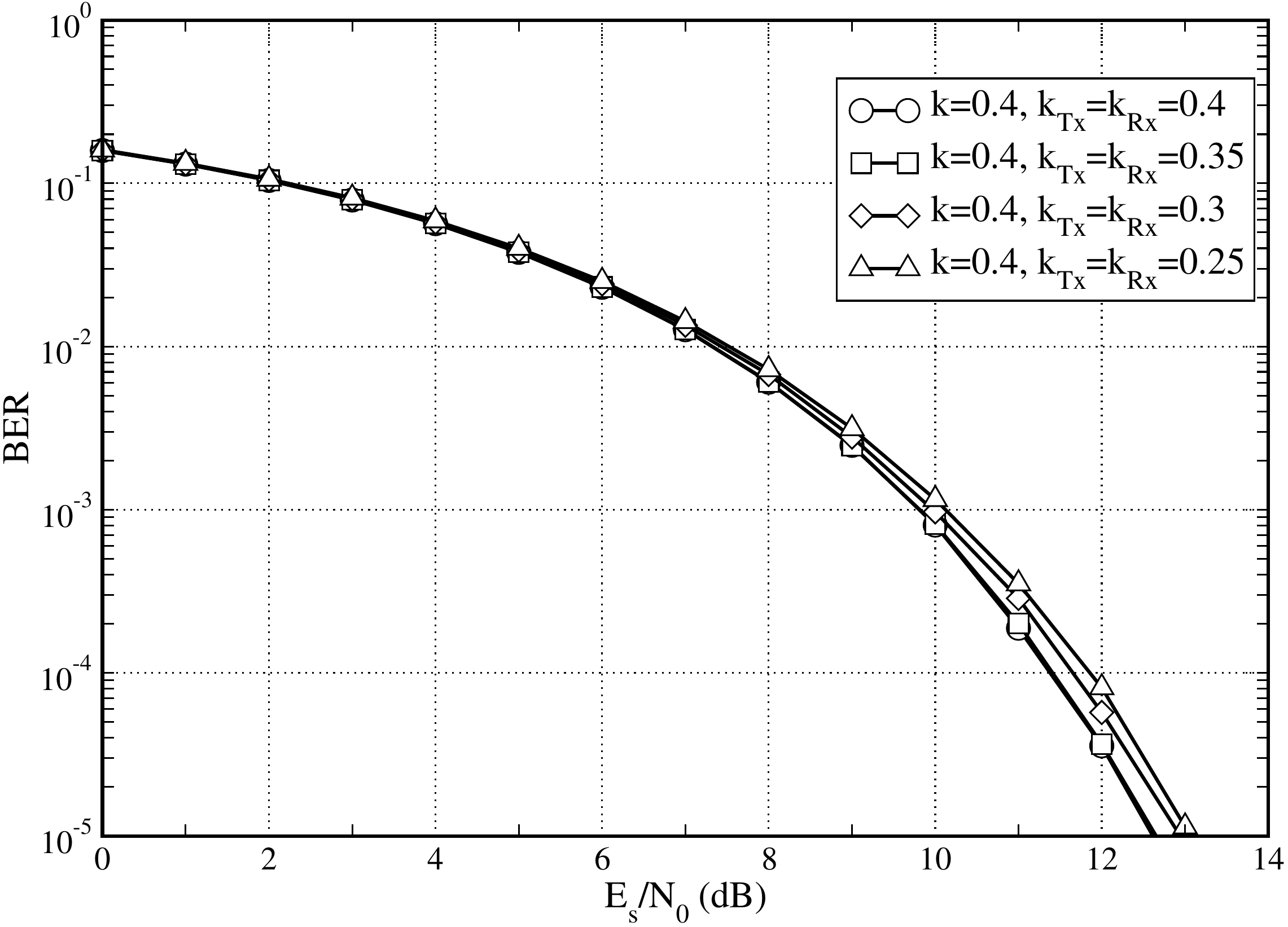}
}\qquad
\subfigure{
\includegraphics[scale=0.35]{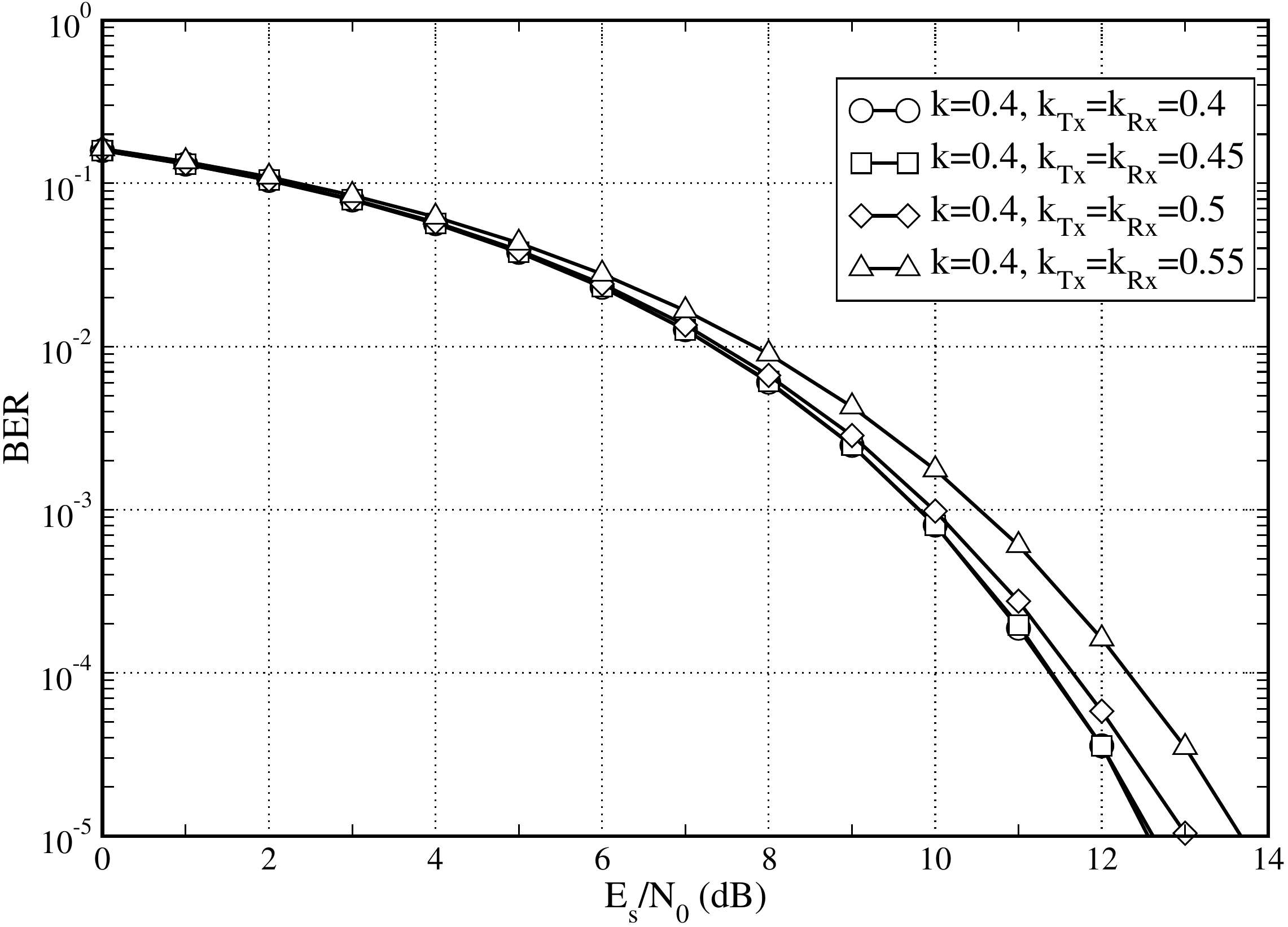}
}
\caption{Sensitivity of bit error rate for 2-ary FSK as a function of a wrongly estimated coupling coefficient at a data rate of 10~kbps.}
\label{ber_frequencyoffset}
\end{figure*}

In Fig.~\ref{ber_frequencyoffset} the influence of a frequency mismatch is studied for 2-ary FSK.  Towards this goal, the coupling coefficient $k$ 
on the channel is fixed to be $k=0.4$.  
At Tx and Rx units, the value of $k$ needs to be estimated, denoted as $k_{\mbox{\footnotesize Tx}}$ and $k_{\mbox{\footnotesize Rx}}$.  
It makes sense to assume $k_{\mbox{\footnotesize Tx}}=k_{\mbox{\footnotesize Rx}}$, because Tx and Rx are able to communicate via a feedback link. 
The resonant frequency $f_0=1$~MHz is known perfectly. 
A small data rate is selected in order to avoid ISI completely. 
Any setting $k\not=k_{\mbox{\footnotesize Tx}}=k_{\mbox{\footnotesize Rx}}$ suffers from a frequency mismatch.  
As observed from Fig.~\ref{ber_frequencyoffset}, 
even for significant offsets no error floor is visible in the illustrated range of BERs. 
Underestimating $k$ is more robust compared to an overestimation.  

\section{Experimental Results}\label{sec7}
\begin{figure}[t]
\centering
\includegraphics[scale=0.35]{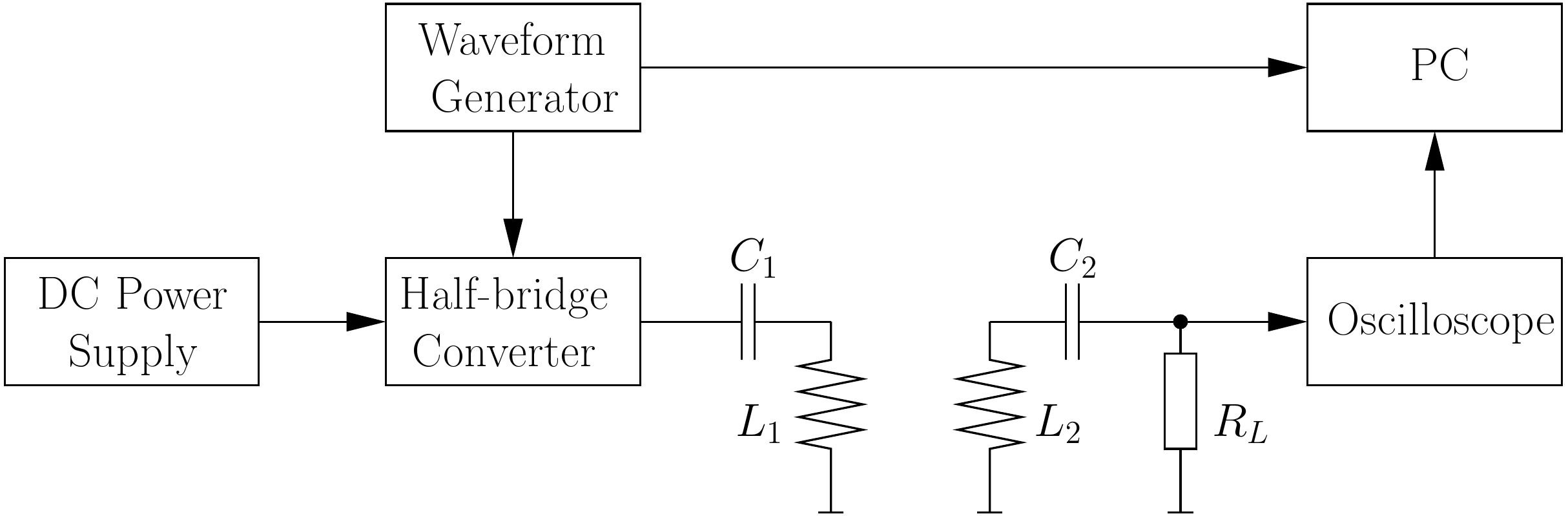}
\caption{Schematic diagram of conducted experiment. The included components are: 
DC power supply, waveform generator, half-bridge converter, Tx and Rx resonant circuits, Ohmic load, digital sampling oscilloscope, Linux PC. 
The PC performs offline processing including receive filtering.}
\label{devicediagram}
\end{figure}

In order to support the simulation results, we have implemented an experimental prototype in the 10~W regime.
The experimental device diagram is shown in Fig.~\ref{devicediagram}. 
The resonant frequency is chosen to be $f_0=1$~MHz. 
The core is a WE~760308111 wireless power charging coil manufactured by W\"urth Elektronik. 
This coil is mounted on a ferrite plate and has an inductance of $\SI{6.3}{\micro \henry} \pm 10~\%$ over a wide frequency range. 
The self resonant frequency is documented as 11~MHz.  
The quality factor is about $Q=64$ at 1~MHz.  
Correspondingly, $R_1=R_2=0.62~\Omega$ at 1~MHz if the same type of coil is used at both sides.
The inductance is compensated by $C_1=C_2=4$~nF.  NP0/C0G class 1 ceramic capacitors are used, 
because they offer high stability and low losses for resonant circuit applications.  Their temperature coefficient is very small: 
The capacitance variation is $\pm 0.3$~\% within the wide temperature range of $\SI{-55}{\degreeCelsius}$ to $\SI{+125}{\degreeCelsius}$.
A half-bridge converter has been implemented based on an IXD~630 ultrafast MOSFET driver in order to generate unipolar RFSK. 
The logic input signal has been generated by function/arbitrary waveform generator Rigol DG5072. 
The driver is capable to deliver an output current of up to $\pm 8$~A in continuous mode.   
The DC supply voltage of the driver is selected to be 12~V. 
The output resistance is about $R_S=0.17~\Omega$ at high state and low state according to the data sheet. 
Mark and space frequencies of the RFSK signal are adjusted as $f_{+}=1.291$~MHz and $f_{-}=0.845$~MHz, respectively. 
The load, for example a rechargeable battery including rectifier, 
is emulated by an equivalent load resistance, which is assumed to be $R_L=10~\Omega$. 
The load will be slowly time-varying in practice, but this effect has no impact on the position of the peak frequencies if $k>k_{\mbox{\footnotesize split}}$.
The entire set of parameters has been used consistently throughout the article.
The data rate scales with the resonant frequency $f_0$. 1~MHz is sufficiently large for reasonable data rates, 
yet sufficiently small concerning losses typically encountered in the 10~W regime. 
$f_0=1$~MHz serves as an example, rather being claimed to be best practice.  
In low-power use cases, a larger resonant frequency is not uncommon.  Vice versa, 
in power electronic applications normally a lower resonant frequency would be chosen, though particularly gallium nitride (GaN) 
technology is shifting borders as well.

\begin{figure}[t]
\centering
\includegraphics[scale=0.24]{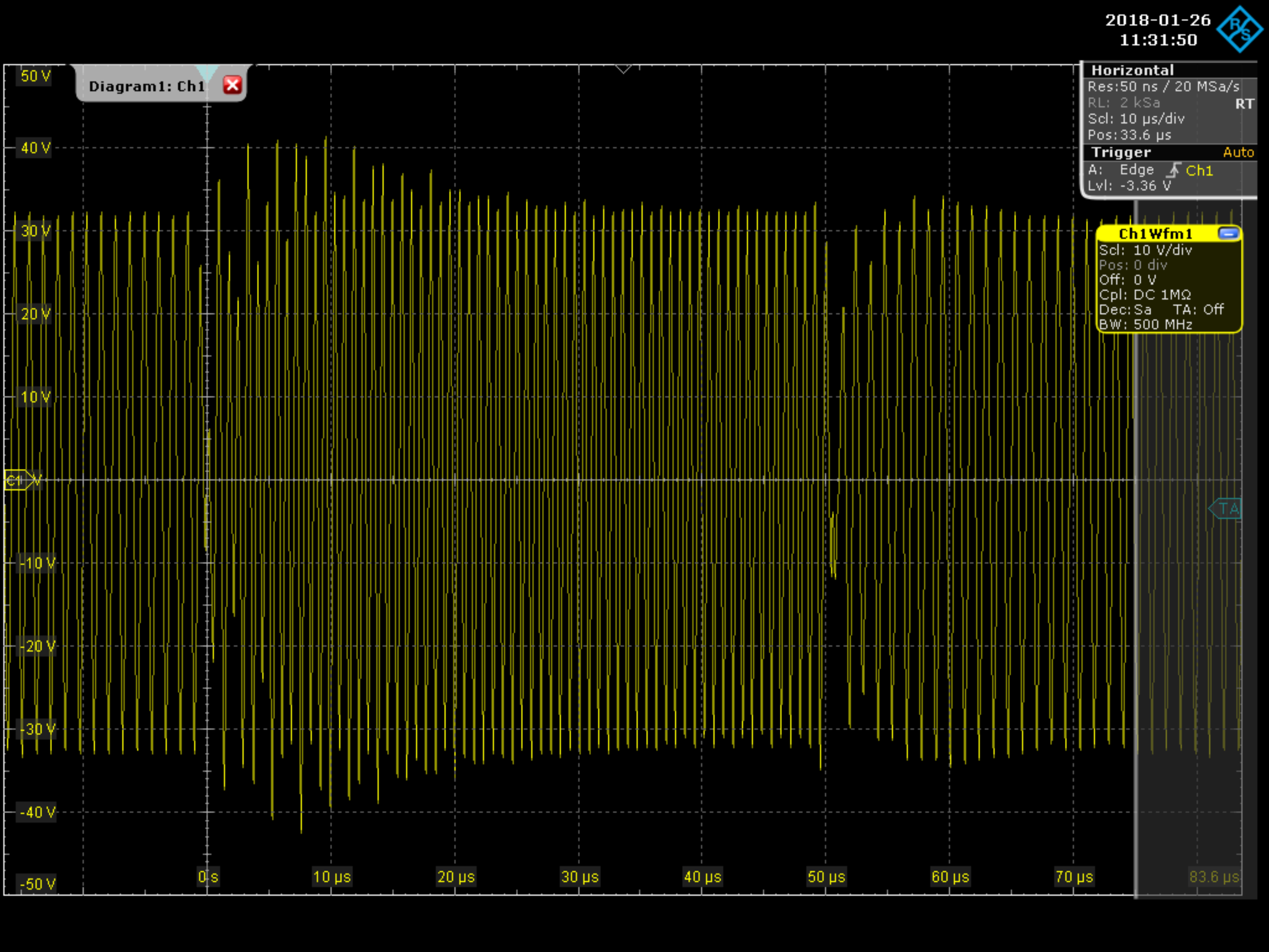}
\caption{Screenshot of the received voltage given a symbol duration of $\SI{50}{\micro s}$ (20~kbps).}
\label{experiment}
\end{figure}

The received waveform across coil $L_2$
has been sampled by a Rohde \& Schwarz RTO~1004 oscilloscope with a sampling rate of 20~MSa/s
providing 8~bits/sample, see Fig.~\ref{experiment}.  Sequences of 80~MSa have been evaluated offline by software.  
In order to study the transient behavior, the data bits are toggled alternatingly.
Due to the frequent symbol transitions, this establishes the worst case.
Transfer function and BER results, however, have been obtained for pseudo-random binary sequences. 
The distance between the coils has been optimized so that the measured magnitude of the transfer function 
displayed on the oscilloscope 
matches the theoretical transfer function shown in Fig.~\ref{fig2}, i.e., the coupling coefficient is about $0.4$ in the experiment.
For the coil layout used in the experiment (44~mm $\pm$~1.5~mm outer diameter, mounted on ferrite plate), 
at $k=0.4$ the coil distance is about 10~mm if the coils are perfectly aligned. 

\begin{figure}[t]
\centering
\includegraphics[scale=0.35]{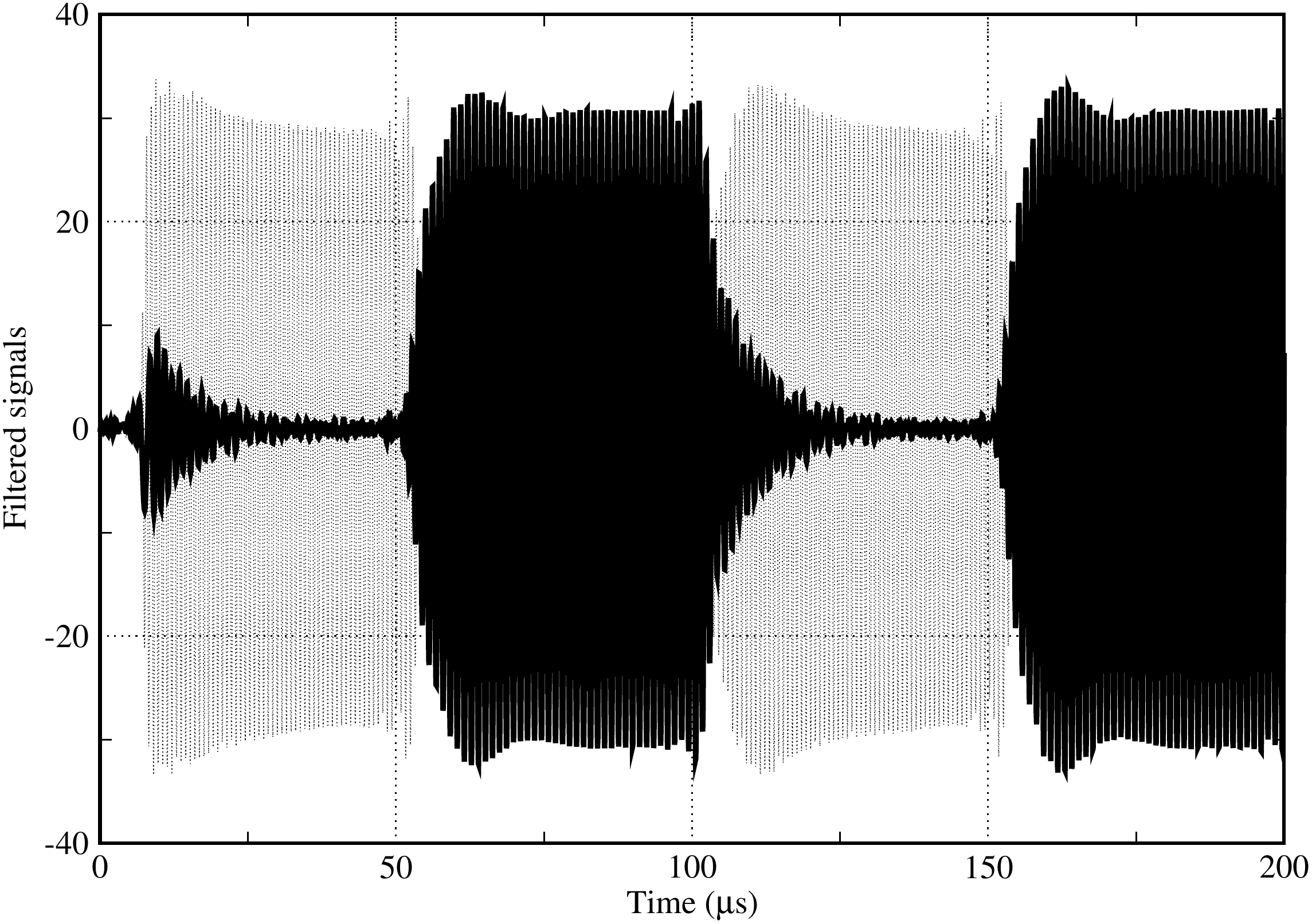}
\caption{Measured received signals after low-pass filtering (black) and band-pass filtering (gray) given a symbol duration of $\SI{50}{\micro s}$ (20~kbps).}
\label{filtered}
\end{figure}

In the numerical results, sensitivity tests are based on a coherent receiver in order to provide a lower bound on the BER.
In the experimental prototype, a low-cost noncoherent receiver consisting of two parallel filters has been used instead: A low-pass filter
with a cut-off frequency $f_0$ (at the end of the pass band) and a band-pass filter with the identical cut-off frequency
(at the start of the pass band).
Both filters have the same attenuation in their pass bands and the same effective noise bandwidth.
They have been designed by Matlab (equiripple design, 1~MHz bandwidth, 30~dB stop band attenuation, 0.4~dB pass band ripple, 291 taps).
The signals at the filter outputs, shown in Fig.~\ref{filtered}, are rectified and then averaged over one symbol duration after removing the cyclic extension.
Whenever the averaged output signal of the low-pass filter
exceeds the one of the band-pass filter, $\hat u[\kappa]=0$ is declared.  Otherwise, $\hat u[\kappa]=1$ is declared.
In the presence of a time-varying $k$, an adjustment of receive filtering is not necessary.
Although the noncoherent receiver algorithm is suboptimal w.r.t.~BER performance, not a single error event has been observed
for data rates up to 100~kbps.
This is due to the fact that in the 10~W prototype distortions are small, compare Fig.~\ref{experiment} with Fig.~\ref{signal},
given the short distances of interest.

The following statements drawn throughout this contribution could be verified by the experiment: 
(i) The impulse response has the expected shape and duration (see Fig.~\ref{experiment}). 
(ii) The efficiency factor is almost independent of the data rate (measured in a range between 2~kbps and 100~kbps). 
(iii) An efficiency of about 81.4~\% could be achieved at $k\approx 0.4$ for alternating data symbols. 
This figure neither includes the power consumption of the MOSFET driver nor the voltage loss of a rectifier, 
in order to provide a fair comparison with the theoretical efficiency of 84.9~\% obtained in Sections~\ref{sec2} and~\ref{sec6}. 
The gap between the experimental results and theoretical predictions is reasonably small. 

\section{Conclusions}\label{sec8}
In inductively coupled resonant circuits the so-called frequency splitting phenomenon occurs at short distances. 
Frequency splitting both affects data transmission and power transmission. 
Given a two-coil series-resonant circuit with an Ohmic load, both peaks are shown to have about the same gain. 
This observation motivates to exploit frequency splitting when designing the modulation scheme.
Towards this goal, 2-ary FSK in conjunction with a cyclic extension is proposed as a low-cost but efficient modulation scheme.  
Whenever the data bit is $u[\kappa]=0$ the primary loop is excited with maximum possible amplitude at the lower peak frequency $f_{-}$, 
otherwise at the higher peak frequency $f_{+}$. 
As opposed to other modulation schemes suffering from a trade-off between high $Q$ (for energy-efficient power transmission) and low $Q$ 
(for wideband data transmission), remarkably for 2-ary FSK there is almost no loss of output power 
when data transmission is performed simultaneously,
as long as the symbol duration is not shorter than about ten times of the length of the impulse response.
Also interesting is the fact that 
the average efficiency during the transient phase, $\overline{\eta_T}$, is about the same as the average efficiency of the steady state, 
$(\eta_{-}+\eta_{+})/2$. 
Consequently, 2-ary FSK is fairly robust in the scenario of interest. 

As an alternative to conventional 2-ary FSK, rectified FSK is proposed and evaluated. 
Bipolar RFSK can be generated by a full-bridge converter, unipolar RFSK by a half-bridge converter. 
Bipolar RFSK, unipolar RFSK and 2-ary FSK have about the same efficiency.
The output power of bipolar RFSK is four times the output power of unipolar RFSK given the same peak input voltage. 
2-ary FSK is in-between.  Consequently, bipolar RFSK causes the highest SNR.  In practice, bipolar RFSK is even more favorable,  
because a switching-type of power converter typically has a higher power efficiency than a linear generator. 
Furthermore, RFSK is easier to implement compared to FSK. 

On a more theoretical side, an equivalent discrete-time channel model has been derived, which simplifies numerical simulations and 
serves as a basis for analysis.  An inspection of the dominant noise terms reveals that the signal-to-noise ratio is frequency flat, 
independently whether noise comes from the transmitter side or receiver side. 
Coupling coefficients have been derived that enable orthogonal signaling. 

According to our investigations, both data and power transmission deteriorate in the presence of a frequency mismatch. 
It is not required to estimate the entire voltage transfer function, neither at the transmitter nor at the receiver.
An estimation of both peak frequencies is sufficient at the transmitter side. 
(To be more precise, it is sufficient to estimate one of the peak frequencies. 
The other one can be determined according to (\ref{approx}).) 
Concerning peak frequency estimation at the transmitter side, several methods exist or are feasible. 
A well-known strategy makes the input impedance real-valued \cite{niu13}. 
Another technique is to measure incident and reflected powers as a function of frequency \cite{sam11}. 
A similar measurement approach but additionally employing a ZigBee feedback link has been published in \cite{kim12}.
An alternative method to frequency sweeping would be the estimation of the impulse response.
Finally, one may target on an estimation of the coupling coefficient $k$, as $k$ is uniquely related to the two peak frequencies.
The proposed low-cost receiver algorithm is matched to the constant resonant frequency, rather than the peak frequencies. 
Hence, peak frequency estimation becomes obsolete at the receiver side. 

\section*{Acknowledgment}
This work has been inspired by the EKSH project cMODEM.  Many thanks to Prof.~Sabah Badri-Hoeher, Kiel University of Applied Sciences,
 Germany, for project acquisition, and to Maximilian Placzek, Kiel University, Germany, for prototype design and fruitful discussions.

\end{document}